\documentclass[%
 reprint,
 nofootinbib,
 amsmath,amssymb,
 aps,
]{revtex4-2}

\usepackage{graphicx}
\usepackage{dcolumn}
\usepackage{bm}
\usepackage{color}
\usepackage[normalem]{ulem}

\newcommand{\stoiMatrix}{S}
\newcommand{\rateF}{j}
\newcommand{\flux}{j}
\newcommand{\Vc}[1]{\boldsymbol{#1}}
\newcommand{\dd}{\mathrm{d}}
\newcommand{\dt}{\dd t}
\newcommand{\defeq}{:=}

\newcommand{\kcoef}{k}
\newcommand{\Transpose}{T}
\newcommand{\Y}{\mathcal{Y}}

\newcommand{\U}{U}
\newcommand{\C}{C}

\newcommand{\Polytope}{\mathcal{P}}
\newcommand{\KL}{\mathcal{D}}
\newcommand{\X}{\mathcal{X}}
\newcommand{\vSpace}{\mathbb{V}}
\newcommand{\R}{R}
\newcommand{\T}{T}

\newcommand{\Gibbs}{\mathcal{G}}
\newcommand{\FM}{G}
\newcommand{\diag}{\mathrm{diag}}
\newcommand{\identityM}{I}
\newcommand{\ProjM}{P}
\newcommand{\ECR}{J}
\newcommand{\Real}{\mathbb{R}}

\newcommand{\nx}{N}
\newcommand{\mx}{M}
\newcommand{\Tan}{\mathcal{T}}
\newcommand{\coTan}{\mathcal{T}^{*}}

\newcommand{\Variety}{\mathcal{V}}
\newcommand{\Img}{\mathrm{Im}}
\newcommand{\Ker}{\mathrm{Ker}}
\newcommand{\Dim}{\mathrm{dim}}
\newcommand{\IncMatrix}{B}

\newcommand{\eqnref}[1]{Eq. \ref{#1}}
\newcommand{\secref}[1]{Sec. \ref{#1}}
\newcommand{\fgref}[1]{Fig. \ref{#1}}

\newcommand{\Volume}{\Omega}
\newcommand{\variety}{variety }
\newcommand{\varieties}{varieties }

\newcommand{\com}[1]{#1}

\begin{document}

\preprint{APS/123-QED}

\title{Kinetic Derivation of the Hessian Geometric Structure\\ in Chemical Reaction Systems}

\author{Tetsuya J. Kobayashi, Dimitri Loutchko, Atsushi Kamimura, and Yuki Sughiyama}
 \homepage{http://research.crmind.net}
\affiliation{Institute of Industrial Science, The University of Tokyo, 4-6-1, Komaba, Meguro-ku, Tokyo 153-8505 Japan}

\
\date{\today}

\begin{abstract}
The theory of chemical kinetics form the basis to describe the dynamics of chemical systems.
Owing to physical and thermodynamic constraints, chemical reaction systems possess various structures, which can be utilized to characterize important physical properties of the systems.
In this work, we reveal the Hessian geometry which underlies chemical reaction systems and demonstrate how it originates from the interplay of stoichiometric and thermodynamic constraints. 
Our derivation is based on kinetics, we assume the law of mass action and characterize the equilibrium states by the detailed balance condition.
The obtained geometric structure is then related to thermodynamics via the Hessian geometry appearing in a pure thermodynamic derivation.
We demonstrate, based on the fact that both equilibrium and complex balanced states form toric varieties, how the Hessian geometric framework can be extended to nonequilibrium complex balanced steady states.
We conclude that Hessian geometry provides a natural framework to capture the thermodynamic aspects of chemical reaction kinetics. 


\end{abstract}

\maketitle


\section{\label{sec:intro}Introduction}
Chemical kinetics constitutes the basis to describe various and complex behaviors of chemically implemented systems such as metabolic networks and intracellular signaling systems \cite{alon2019,mikhailov2017}.
Since the formulation and establishment of the law of mass action by C.M. Guldberg and P. Waage, the theories of chemical kinetics and chemical dynamics have been studied and developed using techniques from various different disciplines \cite{feinberg2019,alon2019,murray2011,epstein1998,beard2008}.

However, the theories of chemical dynamics are not always consistent with thermodynamics. 
Yet, the consideration of thermodynamics is essential for constructing physically and thermodynamically sound kinetic theories.
For example, the combination of the law of mass action and the detailed balance condition leads to a kinetic characterization of the equilibrium state in a way that is consistent with chemical thermodynamics, as shown already in 1901 by Wegscheider \cite{wegscheider1902Z.FuerPhys.Chem.}.
Since the framework of chemical kinetics does not necessarily obey the detailed balance condition, it enables the investigation not only of equilibrium systems but also of a wide range of nonequilibrium reaction systems.
Sparked by the pioneering work of Hill and Schnakenberg \cite{hill2005,hill1966JournalofTheoreticalBiology,schnakenberg1976Rev.Mod.Phys.}, a thermodynamic foundation has been in development for mass action systems out of equilibrium within the last decades \cite{beard2008,polettini2014J.Chem.Phys.,rao2016Phys.Rev.X,rao2018J.Chem.Phys.,rao2018NewJ.Phys.,avanzini2021J.Chem.Phys.,ge2016ChemicalPhysics}, by employing the knowledge from stochastic thermodynamics \cite{schmiedl2007J.Chem.Phys.}.

In addition, the consideration of constraints of physical or thermodynamic origin also introduces additional intriguing structures into chemical kinetics.
Motivated by the work of Horn and Jackson \cite{horn1972Arch.RationalMech.Anal.a}, who extended equilibrium states to complex balanced states, an algebro-geometric structure of chemical reaction systems was discovered and employed to study mass action systems in applied mathematics \cite{craciun2009JournalofSymbolicComputation,craciun2019MBE,craciun2020ArXiv200811468Math}.
Other structures in chemical kinetics are also unveiled by using tools from graph theory, homological algebra, and others \cite{feinberg2019, vanderschaft2013SIAMJ.Appl.Math.,okada2016Phys.Rev.Lett.,mochizuki2015JournalofTheoreticalBiology,fiedler2015Math.MethodsAppl.Sci.,shinar2010Science,araujo2018NatCommun,hirono2021Phys.Rev.Research}.

Thus, clarification of the interrelation of chemical kinetics and thermodynamics can be a fruitful source of new physics and mathematics of chemical reaction systems \cite{polettini2014J.Chem.Phys.,rao2016Phys.Rev.X,rao2018NewJ.Phys.,gopalkrishnan2014SIAMJ.Appl.Dyn.Syst.,craciun2016ArXiv150102860Math,dickenstein2019BullMathBiol,joshi2017SIAMJ.Appl.Dyn.Syst.,otero-muras2017PLOSComputationalBiology}.

Recently, we found that Hessian geometry provides a natural framework for thermodynamics of chemical reaction systems and used the geometric structure to show that several important results, which were thus far derived only from kinetics, are of pure thermodynamic origin \cite{sughiyama2021ArXiv211212403Cond-MatPhysicsphysics}.
Nonetheless, it is important to clarify how the Hessian geometric structure is linked to chemical kinetics because chemical kinetics and thermodynamics were established historically in an interrelated manner and also because the majority of results for chemical reaction systems are based on mass action kinetics rather than thermodynamics.
This is achieved in this paper.

To this end, we find a close connection between the results of equilibrium systems obtained in algebraic geometry \cite{craciun2009JournalofSymbolicComputation} and the thermodynamics which is encoded in the Hessian structure \cite{sughiyama2021ArXiv211212403Cond-MatPhysicsphysics}.
This combination enables us to grasp the geometric structure of the whole state space, which clarifies the dualistic relation between stoichiometric constraints and thermodynamic constrains manifested as the flatness of the respective dual spaces. 
Thereby, we generalize and extend the information geometric framework for chemical reaction networks limited to a single stoichiometric compatibility class \cite{yoshimura2021Phys.Rev.Research}.

We derive the Hessian geometric structure \cite{shima2007}, which purifies and generalizes some aspects of the information geometric one \cite{amari2016}, by starting from the kinetic characterization of the chemical equilibrium state via the law of mass action and the detailed balance condition.
In the derivation, we show that the equilibrium states of a chemical reaction network are described by a toric variety \cite{craciun2009JournalofSymbolicComputation}.
The analytification of the toric variety plays a fundamental role throughout the paper as it constitutes a generalization of the exponential family well-known in statistics \cite{amari1982Ann.Stat.,amari1985Differential-GeometricalMethodsinStatistics,barndorff-nielsen1986Int.Stat.Rev.Rev.Int.Stat.,amari2000,amari2016}.
Building on the theory of exponential families, we extend the dually flat structure from information geometry \cite{amari1989IEEETrans.Inf.Theory,okamoto1991Ann.Stat.,amari2000} to chemical reaction networks.

The toric parameter representation of the equilibrium variety naturally leads to a dual space $\Y$, which is conjugate to the state space $\X$ of molecular concentrations.
The Hessian geometric structure and associated convex potential functions $\varphi(\Vc{x})$ and $\varphi^{*}(\Vc{y})$ appear on the state space $\X$ and its dual $\Y$: in the former, stoichiometric constraints form a linear coordinate system, whereas, in the latter, the equilibrium variety leads to the definition of a dual linear coordinate system.
Since the equilibrium variety is defined by the parameters of the reaction system, which are specified thermodynamically by the environmental variables, the dual space mathematically captures the role of thermal reservoirs attached to the system.
By comparison with the purely thermodynamic derivation in \cite{sughiyama2021ArXiv211212403Cond-MatPhysicsphysics}, the spaces $\X$ and $\Y$ are thermodynamically related to chemical density and chemical potentials.
The convex functions and associated Bregman divergences are mapped to the thermodynamic free energy of the system and the difference of total entropy.
We also show that the Hessian geometric framework can be naturally extended to the nonequilibrium complex balanced states because the both of equilibrium states and the complex balanced states are described by the same toric variety \cite{craciun2009JournalofSymbolicComputation}. 

In the class of equilibrium systems, our thermodynamical results are special instances of the general theory derived from a purely thermodynamic argument.
In this aspect, this paper is supplementary to our accompanying paper \cite{sughiyama2021ArXiv211212403Cond-MatPhysicsphysics}. 
However, owing to the mass action assumption, we find that the equilibrium manifold has the structure of an algebraic variety, which is not true in the general case.
Thereby, we establish a link to the extensive work carried out in real algebraic geometry \cite{craciun2009JournalofSymbolicComputation}.
Owing to this, we show how our results extend to nonequilibrium complex balanced steady-states, because they share the algebraic structure with the equilibrium states.
We expect this link between thermodynamics via Hessian geometry and algebraic geometry to be even more fruitful in the future.

To make the theory and results more accessible to researchers in chemical reaction network theory who are not necessarily familiar with information or Hessian geometry, we clarify several implicit and confusing identifications of different objects in  conventional textbooks of information geometry \cite{amari2016}.

This paper is organized as follows: 
In \secref{sec:CRN}, we introduce a linear coordinate system in the concentration space $\X$ based on the stoichiometric constraints.
In \secref{sec:EV}, we derive that the set of equilibrium states has the structure of a toric variety and present its parametrization. 
The equilibrium variety is used to define a linear coordinate system in the dual space $\Y$, which is also a nonlinear coordinate system of $\X$ space and yields a dual foliation with the stoichiometric constraints.
In \secref{sec:HG}, we clarify the Hessian geometric structure associated with the dual spaces, and introduce its constituents originating from the first derivatives of the potential functions.
In \secref{sec:DG}, we introduce additional properties originating from the second derivatives.
In \secref{sec:TD}, we relate the derived geometric structure to thermodynamics by employing the results in \cite{sughiyama2021ArXiv211212403Cond-MatPhysicsphysics}.
In \secref{sec:CB}, we show how the framework can be passed on to the nonequilibrium complex balanced state.
In the supplementary \secref{sec:ST}, we demonstrate how our work is relevant to the Markov chain model on a graph, which is often used in stochastic thermodynamics.

\section{Chemical Reaction Kinetics and Stoichiometric Space}\label{sec:CRN}
In this work, we consider reversible chemical reaction networks with the stoichiometric matrix given by $\stoiMatrix \in \mathbb{Z}^{N \times M}$, where $N$ and $M$ are the number of chemical species and that of pairs of forward and reverse reactions, respectively.
Let $\Vc{x}\defeq(x_{1},\cdots,x_{N}) \in \X$ be the concentrations of molecular species in a constant reaction volume $\Volume$ involved in the network, whose state space $\X$ is the positive orthant: $\X \defeq \mathbb{R}_{> 0}^{\nx}$.

Let $\Vc{\rateF}^{\pm}(\Vc{x};\Vc{\theta})$ be the rate functions of the forward and reverse reactions. 
The vector $\Vc{\theta}$ represents parameters of $\Vc{\rateF}^{\pm}$ such as reaction rate constants, the detail of which is specified later and represented abstractly at this point.
The reaction rate equation of the system is given by 
\begin{align}
    \frac{\dd \Vc{x}}{\dt}=\stoiMatrix \Vc{\flux}(\Vc{x};\Vc{\theta}),\label{eq:CRN}
\end{align}
where $\Vc{\flux}(\Vc{x};\Vc{\theta})\defeq \Vc{\rateF}^{+}(\Vc{x};\Vc{\theta})-\Vc{\rateF}^{-}(\Vc{x};\Vc{\theta})$ is the total flux \cite{beard2008,feinberg2019}.

Without loss of generality, we can assume that $\Vc{x}~>~0$ because if $x_{i}=0$ holds for any $i$, then we can construct a reduced stoichiometric matrix $\stoiMatrix_{red}$ by eliminating the $i$-th row from $\stoiMatrix$ and a reduced rate function $\Vc{\flux}_{red}(\Vc{x};\Vc{\theta})$ by inserting $x_{i}=0$ to $\Vc{\flux}(\Vc{x};\Vc{\theta})$.

\subsection{Conserved quantities and stoichiometric polytope}
From the rate equation, we can see that, for any column vector $\Vc{u} \in \Ker[\stoiMatrix^{\Transpose}]$, the quantity $\Vc{u}^{\Transpose}\Vc{x}(t)$ is conserved \cite{feinberg2019}. 
Let $\{\Vc{u}_{i}\}_{i\in [1,\cdots,\ell]}$ be a basis of $\Ker[\stoiMatrix^{\Transpose}]$ and $\ell$ be the dimension of $\Ker[\stoiMatrix^{\Transpose}]$.
Note that $\{\Vc{u}_{i}\}$ is generally a non-orthogonal (oblique) basis \footnote{We also emphasize that we do not assume any metric at this point.}.
We define $\U\defeq \left(\Vc{u}_{1},\cdots,\Vc{u}_{\ell} \right)^{\Transpose}$.
From this definition, $\stoiMatrix^{\Transpose}\U^{\Transpose}=0$ and $\U \stoiMatrix =0$ hold.
Then, a vector $\Vc{\eta}=\U \Vc{x}_{0}$ specifies the values of all stoichiometically conserved quantities for the initial state $\Vc{x}_{0}$\footnote{The system can have additional (nonlinear) conserved quantities that are determined by the particular structure of $\Vc{\rateF}(\Vc{x})$.}.
The trajectory $\Vc{x}(t)$ of \eqnref{eq:CRN} starting from $\Vc{x}_{0} \in \X$ at $t=0$ satisfies $\U\Vc{x}(t)=\U\Vc{x}_{0}$.
Thus, we define the stoichiometric polytope (stoichiometric compatibility class)  as (\fgref{fg:space_X} a)
\begin{align}
    \Polytope^{\X}(\Vc{\eta})\defeq\{\Vc{x}|\U\Vc{x}=\U\Vc{x}_{0}=\Vc{\eta}\}. \label{eq:polytope}
\end{align}
To ensure that $\Vc{x}>0$, we determine the domain of $\Vc{\eta}$ appropriately, cf. Eq. \ref{eq:E}.

\subsection{Extent of Chemical Reaction}
\begin{figure}
  \begin{center}
   \includegraphics[width=0.45\textwidth]{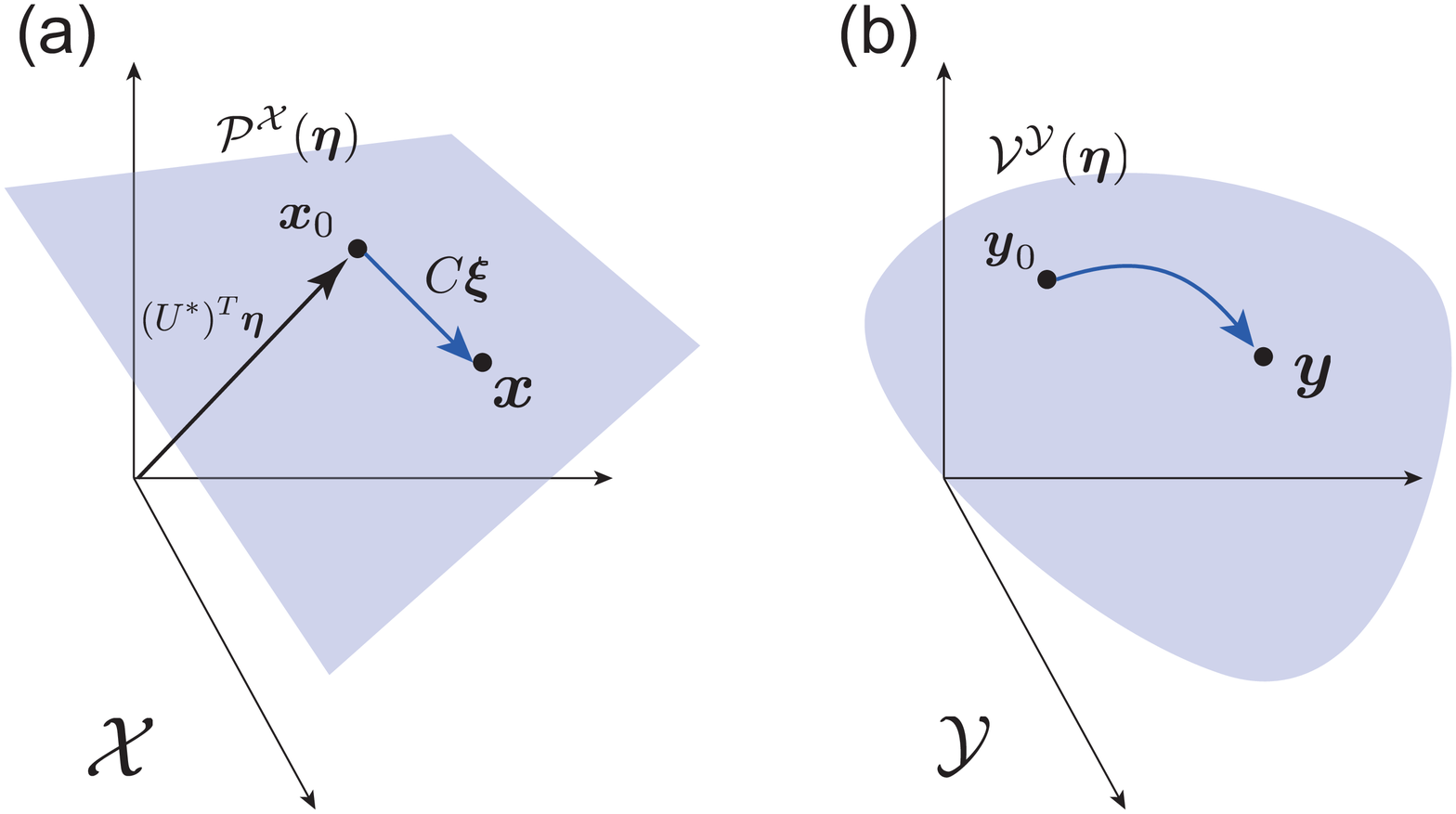}
  \end{center}
  \caption{(a) A linear coordinate system of $\X$ induced by the stoichiometric matrix $\stoiMatrix$. 
  The blue plane represents the stoichiometric polytope $\Polytope^{\X}(\Vc{\eta})$.
  (b) The stoichiometric manifold $\Variety^{\Y}(\Vc{\eta})$ (the blue curved surface) obtained by mapping $\Polytope^{\X}(\Vc{\eta})$ into $\Y$ by the Legendre transformation $\partial \varphi$. }
  \label{fg:space_X}
 \end{figure}

The state of the system $\Vc{x}(t)$ is restricted to the stoichiometric polytope: $\Vc{x}(t) \in \Polytope^{\X}(\Vc{\eta})$. 
By using the extent of chemical reaction $\Vc{\ECR}\defeq \int_{0}^{t} \Vc{\flux}(\Vc{x}(t');\Vc{\theta})\dt' \in \Real^{\mx}$,
we can specify $\Vc{x}(t)$ starting from $\Vc{x}_{0}\in \Polytope^{\X}(\Vc{\eta})$ as 
\begin{align}
\Vc{x}(t) = \Vc{x}_{0} + \stoiMatrix \Vc{\ECR}\in \Polytope^{\X}(\Vc{\eta}).
\end{align}
More generally, because $\Vc{x}-\Vc{x}_{0} \in \Img [\stoiMatrix]$, any point on $\Polytope^{\X}(\Vc{\eta})$ can be specified as
\begin{align}
\Vc{x} = \Vc{x}_{0} + \C \Vc{\xi}\in \Polytope^{\X}(\Vc{\eta}), \label{eq:xrep1}
\end{align}
where $\Vc{\xi} \in \Real^{N-\ell}$ and $\C$ is defined by using $N-\ell$ independent vectors $\{\Vc{c}_{1},\cdots,\Vc{c}_{N-\ell}\}$ that form an oblique basis of $\Img [\stoiMatrix]$.
Note that $N-\ell = \Dim~\Img [\stoiMatrix]$.
Because $\Vc{u}_{i}^{\Transpose}\Vc{c}_{j}=0$ for all $i$ and $j$, the equality $\U \C=0$ holds because of $\U \stoiMatrix =0$ (\fgref{fg:space_X} a).

We define the dual bases $\U^*$ and $\C^*$ of $\U$ and $\C$ such that the orthogonality relations $\U^{*}\U^{\Transpose}=\identityM$, $\C^{\Transpose}\,\C^{*}=\identityM$, and $\U^{*}\C^{*}=0$ are satisfied.
Then, these bases span the following linear spaces: $\left<\U^{\Transpose}\right>=\Ker [\stoiMatrix^{\Transpose}]$, $\left<\C\right>=\Img [\stoiMatrix]$, $\left<(\U^{*})^{\Transpose}\right>=\Img [\stoiMatrix]^{\perp}$ and $\left<\C^{*}\right>=\Ker [\stoiMatrix^{\Transpose}]^{\perp}$, where $\left<A\right>$ denotes the subspace spanned by vectors in $A$ and $^{\perp}$ denotes the orthogonal complement \cite{horn2013}.

\eqnref{eq:xrep1} is not a canonical representation of a given $\Vc{x}$.
In other words, for any $x \in \X$, the coordinates $\Vc{\xi}$ in \eqnref{eq:xrep1} are not uniquely given because they depend on the choice of $\Vc{x}_{0}\in \Polytope^{\X}(\Vc{\eta})$.
To make the representation unique, among all $\Vc{x}_{0}\in \Polytope^{\X}(\Vc{\eta})$, we choose one that satisfies $\Vc{x}_{0}(\Vc{\eta})=(\U^{*})^{\Transpose}\Vc{\eta}$.
Because $\U(\U^{*})^{\Transpose}=\identityM$ holds, $\Vc{\eta}=\U \Vc{x}_{0}(\Vc{\eta})$ is satisfied. 
This gives a unique linear parametrization of $\Vc{x}$ as  
\begin{align}
\Vc{x}(\Vc{\eta},\Vc{\xi}) = (\U^{*})^{\Transpose}\Vc{\eta} + \C \Vc{\xi}, \label{eq:xrep}
\end{align}
where $\Vc{\eta}$ specifies the position of the origin of stoichiometric polytope and $\Vc{\xi}$ is a coordinate on the polytope (\fgref{fg:space_X} a)\footnote{Note that \eqnref{eq:xrep} is obtained without any kinetic information.}.

With this parametrization, the stoichiometric polytope is represented as
\begin{align}
    \Polytope^{\X}(\Vc{\eta})\defeq\{\Vc{x}|\Vc{x}=(\U^{*})^{\Transpose}\Vc{\eta} + \C \Vc{\xi},(\Vc{\eta},\Vc{\xi}) \in \mathcal{E} \}. \label{eq:polytopeparam}
\end{align}
where we define 
\begin{align} \label{eq:E}
    \mathcal{E} \defeq\{(\Vc{\eta},\Vc{\xi})|(\U^{*})^{\Transpose}\Vc{\eta} + \C \Vc{\xi}>\Vc{0}\}, 
\end{align}
to ensure $\Vc{x}>0$.
In the following, we consider only $(\Vc{\eta},\Vc{\xi})\in \mathcal{E}$.

\section{Equilibrium variety and dual coordinate}\label{sec:EV}
The conserved quantities $\Vc{\eta}$ and the extent of chemical reaction $\Vc{\ECR}$, or its variant $\Vc{\xi}$, are commonly used as variables to characterize chemical systems. 
As verified in \eqnref{eq:xrep}, $(\Vc{\eta}, \Vc{\xi})$ is a linear coordinate system of $\X$.
Next, we show how dual coordinates $(\Vc{\eta}^{*}, \Vc{\xi}^{*})$ can be naturally defined if the system is an equilibrium system.

We point out that, to obtain the result, we will use the detailed balance condition together with the kinetic law of mass action to characterize the equilibrium state and the system. 
In our accompanying paper, by contrast, we derive the same result only from a thermodynamic argument without these kinetic assumptions \cite{sughiyama2021ArXiv211212403Cond-MatPhysicsphysics}. 
We use the kinetic assumptions here because these assumptions prevail in chemical reaction network theory and stochastic thermodynamics \cite{ge2016ChemicalPhysics,rao2016Phys.Rev.X,schmiedl2007J.Chem.Phys.}, and also because we want to link these disciplines to the results in \cite{sughiyama2021ArXiv211212403Cond-MatPhysicsphysics}.

\subsection{Equilibrium variety}
The positive equilibrium states of the system (\eqnref{eq:CRN}) are the states that satisfy the detailed balance condition:
\begin{align}
\Variety_{eq}^{\X}(\Vc{\theta})\defeq  \{\Vc{x}> 0|\Vc{\flux}(\Vc{x};\Vc{\theta})=\Vc{0}\}.    
\end{align}
Note that $\Variety_{eq}^{\X}(\Vc{\theta})$ can be empty if a specified $\theta$ admits no equilibrium state.
We define the set of parameters $\Theta_{eq}$ such that 
$\Variety_{eq}^{\X}(\Vc{\theta})$ is not empty if $\Vc{\theta} \in \Theta_{eq}$.

To this end, we additionally assume that $\Vc{\rateF}^{\pm}(\Vc{x};\Vc{\theta})$ satisfy the law of mass action:
\begin{align}
    \rateF^{+}_{m}(\Vc{x};\Vc{\theta}) =\kcoef^{+}_{m} \Vc{x}^{\Vc{\gamma}_{m}^{+}},\qquad \rateF^{-}_{m}(\Vc{x};\Vc{\theta}) =\kcoef^{-}_{m} \Vc{x}^{\Vc{\gamma}_{m}^{-}}, \label{eq:LawOfMassActionInd}
\end{align}
where $\kcoef^{\pm}_{m} \in \mathbb{R}_{>0}$ are the rate constants of the $m$th forward and reverse reactions, respectively.
The integer vectors $\Vc{\gamma}_{m}^{+}, \Vc{\gamma}_{m}^{-} \in \mathbb{Z}^{N}$ specify reactants and products of the $m$th forward reaction. 
Thus, $\Vc{\gamma}_{m}^{-}-\Vc{\gamma}_{m}^{+}=\Vc{s}_{m}$ where $\Vc{s}_{m}$ is the $m$th column vector of the stoichiometric matrix $\stoiMatrix$.
For a pair of vectors $\Vc{x}\in\X$ and $\Vc{\alpha}\in\mathbb{Z}^{N}$, the exponential $\Vc{x}^{\Vc{\alpha}}$ represents the monomial 
\begin{align}
\Vc{x}^{\Vc{\alpha}}:=\prod_{i=1}^{N}x_{i}^{\alpha_{i}}.
\end{align}
We write \eqnref{eq:LawOfMassActionInd} in a vector form as 
\begin{align}
    \Vc{\rateF}^{+}(\Vc{x};\Vc{\theta}) =\Vc{\kcoef}^{+}\circ \Vc{x}^{\Gamma^{+}}, \qquad \Vc{\rateF}^{-}(\Vc{x};\Vc{\theta}) =\Vc{\kcoef}^{-}\circ \Vc{x}^{\Gamma^{-}},
\end{align}
where $\Vc{x}^{\Gamma}:=(\Vc{x}^{\Vc{\gamma}_{1}}, \cdots, \Vc{x}^{\Vc{\gamma}_{M}})^{\Transpose}$ and $\circ$ is the component-wise product of vectors\footnote{If necessary, one may adopt the more general version $\Vc{\rateF}^{\pm}(\Vc{x};\Vc{\theta}) =\Vc{\kcoef}^{\pm}\circ \Vc{h}(\Vc{x})\circ \Vc{x}^{\mp\stoiMatrix^{\Transpose}}$ for $\Vc{h}(\Vc{x})>0$.}. 
Then $\Variety_{eq}^{\X}(\Vc{\theta})$ is given by
\begin{align}
    \Variety_{eq}^{\X}(\Vc{\theta}) \defeq  \left\{\Vc{x}> 0| \Vc{\kcoef}^{+}\circ \Vc{x}^{\Gamma^{+}}=\Vc{\kcoef}^{-}\circ \Vc{x}^{\Gamma^{-}}\right\}.
\end{align}
Now, $\Variety_{eq}^{\X}(\Vc{\theta})$ is an algebraic variety, i.e., the manifold defined as the zeros of algebraic equations\footnote{It should be noted that, hereafter, we use the analytifaction of the variety to work on differential geometric aspects of $\Variety_{eq}^{\X}$. But we abuse the word variety to emphasize the fact that $\Variety_{eq}^{\X}$ is given by algebraic equations (which are derived from the detailed balance condition).}.
Thus, $\Variety_{eq}^{\X}(\Vc{\theta})$ is called an equilibrium variety or equilibrium manifold.
In addition, we note that $\Vc{\kcoef}^{\pm}$ are the parameters of $\Vc{\rateF}^{\pm}$ and thus $\Vc{\theta}=(\Vc{\kcoef}^{\pm})$.

\subsection{Parameter conditions for equilibrium}
Next, we derive the necessary and sufficient condition that the set of parameters $\Theta_{eq}$ must satisfy to have non-empty $\Variety_{eq}(\Vc{\theta})$ for $\theta \in \Theta_{eq}$.
In other words, we characterize the parameter set that admits equilibrium states.
By rearranging $\Vc{\kcoef}^{+}\circ \Vc{x}^{\Gamma^{+}}=\Vc{\kcoef}^{-}\circ \Vc{x}^{\Gamma^{-}}$, we obtain 
\begin{align}
\ln \Vc{K}\defeq \ln \frac{\Vc{\kcoef}^{+}}{\Vc{\kcoef}^{-}}=\stoiMatrix^{\Transpose}\ln \Vc{x}, \label{eq:eq_cond_of_k} 
\end{align}
where $\Gamma^{-}-\Gamma^{+}=\stoiMatrix$ was used.
From the Fredholm alternative, \eqnref{eq:eq_cond_of_k} has a solution if and only if $\ln \Vc{K} \in \Img [\stoiMatrix^{\Transpose}]$.
The condition $\ln \Vc{K} \in \Img [\stoiMatrix^{\Transpose}]$ is an abstract representation of the Wegscheider condition \cite{wegscheider1902Z.FuerPhys.Chem.,rao2016Phys.Rev.X}.
Thus, we can represent $\Theta_{eq}$ as
\begin{align}
    \Theta_{eq} =\left\{\Vc{\theta}=(\Vc{\kcoef}^{\pm})|\Vc{K}=\frac{\Vc{\kcoef}^{+}}{\Vc{\kcoef}^{-}},\, \ln \Vc{K} \in \Img [\stoiMatrix^{\Transpose}]\right\}. \label{eq:K}
\end{align}
Hereby, $\Vc{K}$ is the vector of equilibrium constants. 
Thus, this representation means that, among all parameters in $\Theta_{eq}$, only the equilibrium constants $\Vc{K}$ are relevant for the existence of equilibrium states. 
This is natural because the equilibrium state of a system should be characterized statically without specifying any kinetic information of the system.

\subsection{Toric parameterization of the equilibrium variety}
For a given $\Vc{K}\in \Theta_{eq}$, there exists a particular solution $\tilde{\Vc{x}}_{eq}$ of \eqnref{eq:eq_cond_of_k}.
Then the equilibrium \variety $\Variety_{eq}^{\X}(\Vc{K})$, i.e. the set of $\Vc{x}$ that satisfies \eqnref{eq:eq_cond_of_k}, can be represented as
\begin{align}
    \Variety_{eq}^{\X}(\Vc{K})&=\left\{\Vc{x}|\ln \Vc{x}=\ln \tilde{\Vc{x}}_{eq} + \U^{\Transpose}\Vc{\eta}^{*}, \Vc{\eta}^{*}\in \mathbb{R}^{\ell}\right\}, \label{eq:VarEq}
\end{align}
because $U$ is a basis of $\Ker [\stoiMatrix^{\Transpose}]$.
This representation is known as the affine toric parametrization of the \variety $\Variety_{eq}^{\X}(\Vc{K})$.
In algebraic statistics, $\U^{\Transpose}$ is also called the design matrix of the toric variety.
A toric variety is characterized as being generated by a toric ideal, i.e., a prime binomial ideal in the coordinate ring of the ambient space $\mathbb{R}^N \supset \X$ \cite{sottile2008arXiv:math/0212044,eisenbud1996DukeMath.J.,cox2011}.
Because the detailed balance condition is nothing but a set of binomial equations, a toric variety is a natural representation of the equilibrium states.

\begin{figure}
  \begin{center}
   \includegraphics[width=0.45\textwidth]{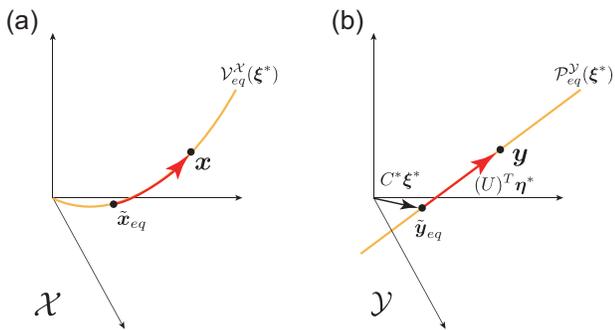}
  \end{center}
  \caption{(a) A curved manifold in $\X$ defined by the equilibrium \variety $\Variety_{eq}^{\X}(\Vc{\xi}^{*})$ (orange curve). $\Variety_{eq}^{\X}(\Vc{\xi}^{*})$ is generally high-dimensional, but in this figure, it is one-dimensional because of the limitation to 3 dimensional space for visualization. (b) the equilibrium variety shown in $\Y$ space. In $\Y$ space, it is a flat subspace. The linear coordinate system induced by the equilibrium variety is also shown. }
  \label{fg:space_Y}
 \end{figure}

From \eqnref{eq:VarEq}, we see that $\Vc{\eta}^{*}$ works as a coordinate of the variety $\Variety_{eq}^{\X}(\Vc{K})$.
However, similarly to the case of $\Vc{x}_{0}$, the actual value of $\Vc{\eta}^{*}$ depends on the choice of $\tilde{\Vc{x}}_{eq}$, which is not uniquely specified because it is just a particular solution.
Among all $\tilde{\Vc{x}}_{eq}$ satisfying $\ln \Vc{K}=\stoiMatrix^{\Transpose}\ln\tilde{\Vc{x}}_{eq}$, we choose one such that $(\ln\tilde{\Vc{x}}_{eq}+\hat{\Vc{y}}) \in \Ker [\stoiMatrix^{\Transpose}]^{\perp}$ where $\hat{\Vc{y}}$ determines a reference point and can be associated with the standard chemical potential, as we will see later.
Because $\left<\C^{*}\right>=\Ker [\stoiMatrix^{\Transpose}]^{\perp}$, we can 
write $\ln\tilde{\Vc{x}}_{eq}=\C^{*}\Vc{\xi}^{*}-\hat{\Vc{y}}$.
Since the equation
\begin{align}
\ln \Vc{K}=\stoiMatrix^{\Transpose}\left(\C^{*}\Vc{\xi}^{*}- \hat{\Vc{y}}\right)\label{eq:lnK_xi}
\end{align}
uniquely determines $\Vc{\xi}^{*}$ if $\hat{\Vc{y}}$ is fixed and $\ln \Vc{K}\in \Img \stoiMatrix^{\Transpose}$, we can use $\Vc{\xi}^{*}$ instead of $\Vc{K}$ to specify the equilibrium \variety, which we denote by $\Variety_{eq}^{\X}(\Vc{\xi}^{*})$ from now on (\fgref{fg:space_Y} a):
\begin{align}
    \Variety_{eq}^{\X}(\Vc{\xi}^{*})=\left\{\Vc{x}|\ln \Vc{x}=\C^{*}\Vc{\xi}^{*} + \U^{\Transpose}\Vc{\eta}^{*}-\hat{\Vc{y}}\right\}. \label{eq:VarEqXi}
\end{align}

Now we introduce the space $\Y$ obtained by the nonlinear transformation 
\begin{align}
\Vc{y}=\ln \Vc{x}+\hat{\Vc{y}},\label{eq:Cojtransform}
\end{align}
i.e., 
$\Y \defeq \hat{\Vc{y}} + \ln \X= \Real^{N}$, any point on $y \in \Y$ can be linearly parametrized as
\begin{align}
    \Vc{y}(\Vc{\eta}^{*},\Vc{\xi}^{*})=\U^{\Transpose}\Vc{\eta}^{*} + \C^{*} \Vc{\xi}^{*}.
\end{align}
Thus, $(\Vc{\eta}^{*},\Vc{\xi}^{*})$ is a linear coordinate system of $\Y$ (\fgref{fg:space_Y} b).
By the reverse transformation of \eqnref{eq:Cojtransform}, any $\Vc{x} \in \X$ can also be parametrized nonlinearly as 
\begin{align}
    \Vc{x}(\Vc{\eta}^{*},\Vc{\xi}^{*})=\exp\left[ \U^{\Transpose}\Vc{\eta}^{*} + \C^{*} \Vc{\xi}^{*}-\hat{\Vc{y}}\right].
\end{align}
As the form of the transformation in \eqnref{eq:Cojtransform} implies, $\Y$ is the space of chemical potentials, which is thermodynamically conjugate to the concentration space $\X$ \cite{callen1985}.
Moreover, $\Vc{x}(\Vc{\eta}^{*},\Vc{\xi}^{*})$ and $\Vc{y}(\Vc{\eta},\Vc{\xi})$ are Legendre dual as we will show in \secref{sec:HG}.

Note that the real algebraic geometry of toric varieties has been employed in computational and algebraic statistics to handle exponential families with constraints \cite{pachter2005}.

\subsection{Foliation and mixed coordinate}
Finally, we show that $(\Vc{\eta},\Vc{\xi}^{*})$, a mixture of the two coordinate systems $(\Vc{\eta},\Vc{\xi})$ and $(\Vc{\eta}^{*},\Vc{\xi}^{*})$, can also work as a nonlinear coordinate system for $\X$. 

\begin{figure}
  \begin{center}
   \includegraphics[width=0.45\textwidth]{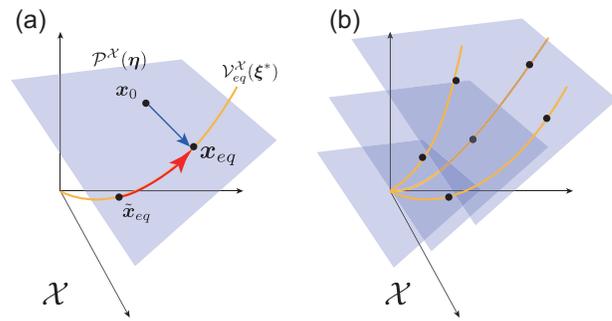}
  \end{center}
  \caption{(a) Intersection of the stoichiometric polytope $\Polytope^{\X}(\Vc{\eta})$ (blue hyperplane) and the equilibrium \variety $\Variety_{eq}^{\X}(\Vc{\xi}^{*})$ (orange curve). 
  (b) Foliation (nonlinear coordinate) formed by the the stoichiometric polytopes (blue hyperplanes) and the equilibrium \varieties (orange curves). }
  \label{fg:space_EQ}
 \end{figure}

For a specific value of kinetic parameters satisfying $\Vc{K} \in \Theta_{eq}$, the \variety $\Variety_{eq}^{\X}(\Vc{K})=\Variety_{eq}^{\X}(\Vc{\xi}^{*})$ specifies the set of equilibrium states for the given parameter value. 
Also, for any initial state $\Vc{x}_{0}$ of the system, its time evolution $\Vc{x}(t)$ is constrained to the stoichiometric polytope $\Polytope^{\X}(\Vc{x}_{0})=\Polytope^{\X}(\Vc{\eta})$.
Thus, the reachable equilibrium point must lie in the intersection (see \fgref{fg:space_EQ} a)
\begin{align}
    \Vc{x}_{eq} \in \Polytope^{\X}(\Vc{\eta}) \cap \Variety_{eq}^{\X}(\Vc{\xi}^{*}). \label{eq:Birch_point}
\end{align}
Because $\Polytope^{\X}(\Vc{\eta})$ and $\Variety_{eq}^{\X}(\Vc{\xi}^{*})$ are characterized by the same structural matrix $\U$, cf. \eqnref{eq:polytope} and \eqnref{eq:VarEqXi}, their intersection is assured to be unique and transversal by Birch's theorem for the exponential family in statistics \cite{pachter2005}.

The uniqueness enables us to specify $\Vc{x}_{eq}$ by $\Vc{\eta}$ and $\Vc{\xi}^{*}$ as $\Vc{x}_{eq}(\Vc{\eta},\Vc{\xi}^{*})$.
Because both $\Polytope(\Vc{\eta})$ and $\Variety_{eq}(\Vc{\xi}^{*})$ can cover the whole state space $\X$ by changing $\Vc{\eta}$ and $\Vc{\xi}^{*}$, respectively, they form a foliation of $\X$ (\fgref{fg:space_EQ} b). 
In other words, $(\Vc{\eta},\Vc{\xi}^{*})$ works as a nonlinear coordinate system of $\X$.
Physically, this means that any equilibrium state $\Vc{x}_{eq}$ can be characterized by the stoichiometric polytope $\Polytope^{\X}(\Vc{\eta})$ and the \variety $\Variety_{eq}^{\X}(\Vc{\xi}^{*})$, each of which has the corresponding value of $\Vc{\eta}$ and $\Vc{\xi}^{*}$ explicitly given by 
\begin{align}
\Vc{\eta}=\U\Vc{x}_{eq}, \qquad \Vc{\xi}^{*}=\C^{\Transpose}(\ln \Vc{x}_{eq}+\hat{\Vc{y}}). \label{eq:eta_xi_x_eq}
\end{align}
This mixed coordinate system is often used in information geometry \cite{amari2016} and the existence of the analogous coordinate system for chemical reaction networks emphasizes their information geometric properties.

\section{Hessian geometric structure of equilibrium chemical reaction systems}\label{sec:HG}
In the previous section, we have introduced the dual coordinates and their mixture for equilibrium chemical reaction systems starting from a conventional chemical kinetics formulation.
Here, we deductively clarify their Hessian  geometric structure \cite{amari2016,shima2007,sughiyama2021ArXiv211212403Cond-MatPhysicsphysics}.
 
\begin{figure}
  \begin{center}
   \includegraphics[width=0.45\textwidth]{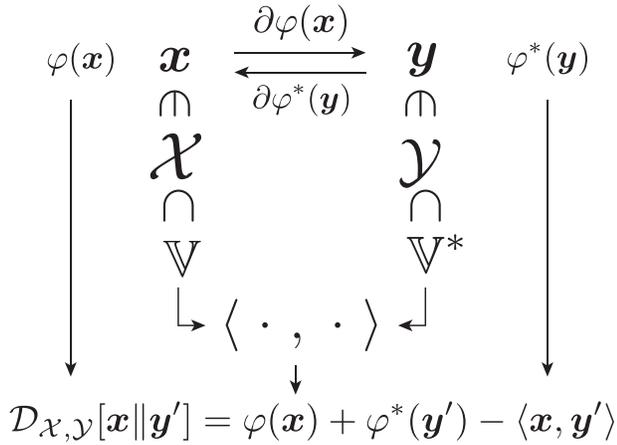}
  \end{center}
  \caption{Relations between $\X$ and $\Y$, $\mathbb{V}$ and $\mathbb{V}^{*}$, $\varphi(\Vc{x})$ and $\varphi(\Vc{x})^{*}$, and $\partial\varphi(\Vc{x})$ and $\partial \varphi(\Vc{x})^{*}$. The Bregman divergence $\KL$ is induced by using all the constituents.}
  \label{fg:duality}
 \end{figure}
 
\subsection{Dual flat state space}
We regard the state space $\X$ as a subspace embedded in the $N$ dimensional vector space  $\vSpace=\Real^{N}$. 
Let $\Y=\vSpace^{*}$ be the dual vector space of $\vSpace$ and $\left<\cdot, \cdot\right>$ be the bilinear form defined on $\vSpace \times \vSpace^{*}$ (\fgref{fg:duality})\footnote{Note that the result in this section is obtained without assuming any inner product structure.}.
On $\X$ and $\Y$, we define the following two strongly convex potential functions, cf. \fgref{fg:duality}:
\begin{align}
    \varphi(\Vc{x}) &\defeq \left[\ln\Vc{x} - \hat{\Vc{y}} - \Vc{1}\right]^{\Transpose}\Vc{x}, \label{eq:potential_function}\\
    \varphi^{*}(\Vc{y}) &\defeq \Vc{1}^{\Transpose}e^{\Vc{y}+\hat{\Vc{y}}}\label{eq:potential_function_dual}.
\end{align}
This yields the one-to-one Legendre duality between $\Vc{x}\in \X$ and $\Vc{y}\in \Y$ as
\begin{align}
\Vc{y}&=\partial\varphi(\Vc{x}):=\left\{\frac{\partial \varphi(\Vc{x})}{\partial\Vc{x}}\right\}=\ln \Vc{x} -\hat{\Vc{y}},\\
\Vc{x}&=\partial\varphi^{*}(\Vc{y}):=\left\{\frac{\partial \varphi^{*}(\Vc{y})}{\partial\Vc{y}}\right\}=e^{\Vc{y}+\hat{\Vc{y}}},
\end{align}
such that $\varphi(\Vc{x})$ and $\varphi^{*}(\Vc{y})$ are also dual satisfying
\begin{align}
\varphi(\Vc{x})+\varphi^{*}(\Vc{y}) -\left<\Vc{x},\Vc{y}\right>=0.
\end{align}
Thus, the pair $(\X,\Y)$ is equipped with two dualities: one is linear algebraic duality and the other is nonlinear Legendre duality induced by the convex function $\varphi(\Vc{x})$.
This is a Hessian structure \cite{shima2007}, which is a mathematical basis underlying information geometry \cite{amari2016}.

We use the following notation: a pair $(\Vc{x},\Vc{y})$ is always treated as two Legendre dual coordinates (\fgref{fg:duality}).
Another pair with the same decoration, e.g., $(\Vc{x}',\Vc{y}')$, will be treated as an another, generally distinct, Legendre dual pair.

\subsection{Duality in subspaces}
This duality is inherited by the linear coordinate systems, which were derived by a chemical kinetic argument (\fgref{fg:space_X} a and \fgref{fg:space_Y} b):
\begin{align}
\Vc{x}(\Vc{\eta},\Vc{\xi}) &= (\U^{*})^{\Transpose}\Vc{\eta} + \C \Vc{\xi},\\
\Vc{y}(\Vc{\eta}^{*},\Vc{\xi}^{*})&=\U^{\Transpose}\Vc{\eta}^{*} + \C^{*} \Vc{\xi}^{*}.
\end{align}
In particular, we have the following partial Legendre duality between $\Vc{\xi}$ and $\Vc{\xi}^{*}$ and $\Vc{\eta}$ and $\Vc{\eta}^{*}$
\begin{align}
    \partial_{\Vc{\eta}}\varphi(\Vc{x}(\Vc{\eta},\Vc{\xi}))&=\U^{*}\Vc{y}=\Vc{\eta}^{*},\\    \partial_{\Vc{\xi}}\varphi(\Vc{x}(\Vc{\eta},\Vc{\xi}))&=\C^{\Transpose}\Vc{y}=\Vc{\xi}^{*},\\
    \partial_{\Vc{\eta}^{*}}\varphi^{*}(\Vc{y}(\Vc{\eta}^{*},\Vc{\xi}^{*}))&=\U\Vc{x}=\Vc{\eta},\\
   \partial_{\Vc{\xi}^{*}}\varphi^{*}(\Vc{y}(\Vc{\eta}^{*},\Vc{\xi}^{*}))&=(\C^{*})^{\Transpose}\Vc{x}=\Vc{\xi},
\end{align}
where $\partial_{\Vc{\eta}}$ is Legendre transform with respect to $\Vc{\eta}$.
We also have
\begin{align}
    \varphi(\Vc{\eta},\Vc{\xi}) + \varphi^{*}(\Vc{\eta}^{*},\Vc{\xi}^{*})=\left<\Vc{\eta},\Vc{\eta}^{*}\right>+\left<\Vc{\xi},\Vc{\xi}^{*}\right>.
\end{align}
This means that the linear coordinate systems defined on $\X$ and $\Y$ are preserved under the additional structure of Legendre duality.
Such $\X$ and $\Y$ are called dually flat spaces in information and Hessian geometry\cite{amari2016,shima2007}.

\subsection{Bregman divergence}
Using the Legendre dual potential functions, the Bregman divergence between two points $\Vc{x}$ and $\Vc{x}'$ is defined as \cite{bregman1967USSRComputationalMathematicsandMathematicalPhysics,amari2016}:
\begin{align}
    \KL_{\X}[\Vc{x}\|\Vc{x}']&\defeq \varphi(\Vc{x})-\varphi({\Vc{x}'})-\left<\partial\varphi(\Vc{x}'), \Vc{x}-\Vc{x}'\right>.
\end{align}
Due to the convexity of $\varphi(\Vc{x})$, the function $\KL[\Vc{x}'\|\Vc{x}]$ is nonnegative and measures the extent of convexity as the deviation of $\varphi(\Vc{x})$ from its linear extrapolation $\varphi({\Vc{x}'})+\left<\partial\varphi(\Vc{x}'), \Vc{x}-\Vc{x}'\right>$, evaluated at $\Vc{x}'$.
The minimum of $\KL_{\X}[\Vc{x}\|\Vc{x}']$ is $0$, which is achieved if and only if $\Vc{x}=\Vc{x}'$.
A direct computation, using \eqnref{eq:potential_function} and \eqnref{eq:potential_function_dual}, gives
\begin{align}
    \KL_{\X}[\Vc{x}\|\Vc{x}']&=\left(\ln\frac{\Vc{x}}{\Vc{x}'}\right)^{\Transpose}\Vc{x}-\Vc{1}^{\Transpose}(\Vc{x}-\Vc{x}').
\end{align}
This indicates that $\KL[\Vc{x}\|\Vc{x}']$ is reduced to the generalized Kullback-Leibler divergence for positive measures on a discrete space for the specific functional form of $\varphi(\Vc{x})$ that originates from the law of mass action and the detailed balance condition.

Similarly, for $\Vc{y}$ and $\Vc{y}'$, the dual Bregman divergence is given by
\begin{align}
\KL_{\Y}[\Vc{y}\|\Vc{y}']&\defeq \varphi^{*}(\Vc{y})-\varphi^{*}({\Vc{y}'})-\left<\partial\varphi^{*}(\Vc{y}'), \Vc{y}-\Vc{y}'\right>.
\end{align}

If $\Vc{y}$ and $\Vc{y}'$ are Legendre dual to $\Vc{x}$ and $\Vc{x}'$, respectively, the Bregman divergences satisfy the symmetry property
\begin{align}
\KL_{\X}[\Vc{x}\|\Vc{x}']=\varphi(\Vc{x})+\varphi^{*}(\Vc{y}')-\left<\Vc{x},\Vc{y}'\right>=\KL_{\Y}[\Vc{y}'\|\Vc{y}],
\end{align}
where $\varphi(\Vc{x}')+\left<\partial\varphi(\Vc{x}'),\Vc{x}' \right>= -\varphi^{*}(\Vc{y}')$ was used. 
Thus we are led to define $\KL_{\X,\Y}[\Vc{x}\|\Vc{y}']$ as (\fgref{fg:duality})
\begin{align}
    \KL_{\X,\Y}[\Vc{x}\|\Vc{y}']\defeq \varphi(\Vc{x})+\varphi^{*}(\Vc{y}')-\left<\Vc{x},\Vc{y}'\right>.
\end{align}
Note that $\KL_{\X}[\Vc{x}\|\Vc{x}']$, $\KL_{\Y}[\Vc{y}'\|\Vc{y}]$, and $\KL_{\X,\Y}[\Vc{x}\|\Vc{y}']$ are just different representations of the same geometric quantity because $\Vc{x}$ and $\Vc{y}$ as well as $\Vc{x}'$ and $\Vc{y}'$ are in one-to-one correspondence by the Legendre transformation.
In the following, we abbreviate $\KL_{\X}[\Vc{x}\|\Vc{x}']$, $\KL_{\Y}[\Vc{y}'\|\Vc{y}]$, and $\KL_{\X,\Y}[\Vc{x}\|\Vc{y}']$ as $\KL[\Vc{x}\|\Vc{x}']$, $\KL[\Vc{y}'\|\Vc{y}]$, and $\KL[\Vc{x}\|\Vc{y}']$, and switch between the three equivalent notations depending on the purpose.
We emphasize that 
\begin{align}
\Vc{x}' &= \arg \min_{\Vc{x}}\KL_{\X}[\Vc{x}\|\Vc{x}'] =\arg \min_{\Vc{x}}\KL_{\X,\Y}[\Vc{x}\|\Vc{y}'].
\end{align}

As we will show later, the relation between the potential function $\varphi$ and the Bregman divergence is the mathematical reason why the Kullback-Leibler divergence appears as the difference of the total entropy in equilibrium systems.

\subsection{Dual orthogonality}
We demonstrate that the Bregman divergence and the mixed coordinate representation play a central role when determining the equilibrium state of the system.

\begin{figure}
  \begin{center}
   \includegraphics[width=0.45\textwidth]{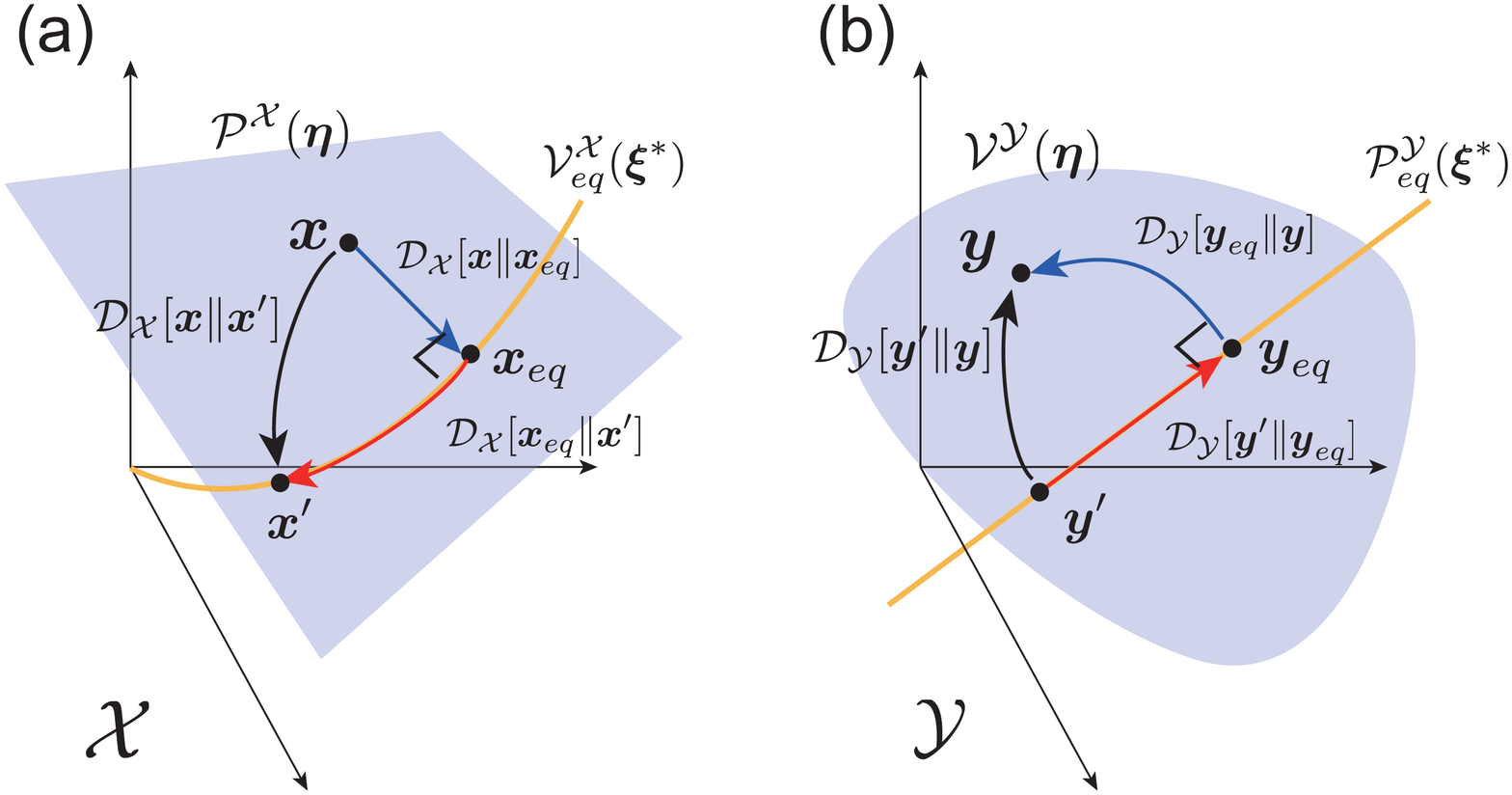}
  \end{center}
  \caption{(a) Graphic explanation of the generalized Pythagorean theorem and orthogonality relation on $\X$ space (\eqnref{eq:PythagoreanX}). The equilibrium point $\Vc{x}_{eq}$ is characterized by either projection of $\Vc{x}$ onto $\Variety_{eq}^{\X}(\Vc{\xi}^{*})$ along the flat subspace  $\Polytope^{\X}(\Vc{\eta})$ or that of $\Vc{x}'$ on the $\Polytope^{\X}(\Vc{\eta})$ along the curved manifold  $\Variety_{eq}^{\X}(\Vc{\xi}^{*})$ (\eqnref{eq:variational_x_2}).
  (b) The same Pythagorean theorem and orthogonarity relation in $\Y$ space (\eqnref{eq:PythagoreanY}). The equilibrium point $\Vc{y}_{eq}$ is characterized by either projection of $\Vc{y}$ onto $\Polytope_{eq}^{\Y}(\Vc{\xi}^{*})$ along the curved manifold  $\Variety^{\Y}(\Vc{\eta})$ or that of $\Vc{y}'$ on the $\Variety^{\Y}(\Vc{\eta})$ along the flat manifold  $\Polytope_{eq}^{\Y}(\Vc{\xi}^{*})$ (\eqnref{eq:variational_y_2})}
  \label{fg:orthogonality}
 \end{figure}

Consider a the chemical reaction system (\eqnref{eq:CRN}) with an equilibrium parameter $\Vc{K} \in \Theta_{eq}$ and initial state $\Vc{x}_{0}$.
Let $\Vc{x}_{0}$ be in stoichiometric polytope $\Polytope^{\X}(\Vc{\eta})$ and let the equilibrium \variety corresponding to $\Vc{K}$ be $\Variety_{eq}^{\X}(\Vc{\xi}^{*})$.
This correspondence is given explicitly in \eqnref{eq:eta_xi_x_eq} and \eqnref{eq:lnK_xi}.
Then, the equilibrium state that the system should converge to is determined by $\Vc{x}_{eq}(\Vc{\eta},\Vc{\xi}^{*})\in \Polytope^{\X}(\Vc{\eta}) \cap \Variety_{eq}^{\X}(\Vc{\xi^{*}})$.
Any point on $\Polytope^{\X}(\Vc{\eta})$, including the initial state $\Vc{x}_{0}$, can be represented uniquely as $\Vc{x}_{p}(\Vc{\eta},\Vc{\xi}_{p}) \in \Polytope^{\X}(\Vc{\eta})$. 
Similarly, any point on $\Variety_{eq}(\Vc{\xi}^{*})$ can be written as $\Vc{x}_{q}^{\X}(\Vc{\eta}_{q}^{*},\Vc{\xi}^{*}) \in \Variety_{eq}^{\X}(\Vc{\xi}^{*})$.
Then, from the definition of $\Polytope^{\X}(\Vc{\eta})$ and $\Variety_{eq}^{\X}(\Vc{\xi}^{*})$, the relations
\begin{align}
    \Vc{x}_{p}(\Vc{\eta},\Vc{\xi}_{p})-\Vc{x}_{eq}(\Vc{\eta},\Vc{\xi})&=\C(\Vc{\xi}_{p}-\Vc{\xi}),\\    
    \Vc{y}_{q}(\Vc{\eta}_{q}^{*},\Vc{\xi}^{*})-\Vc{y}_{eq}(\Vc{\eta}^{*},\Vc{\xi}^{*})&=\U^{\Transpose}(\Vc{\eta}_{q}^{*}-\Vc{\eta}^{*}),
\end{align}
hold.
Here, $\Vc{y}_{q}$ and $\Vc{y}_{eq}$ are the Legendre duals of $\Vc{x}_{q}$ and $\Vc{x}_{eq}$. 
This yields the orthogonality
\begin{align}
    \left<\Vc{x}_{p}(\Vc{\eta},\Vc{\xi}_{p})-\Vc{x}_{eq}(\Vc{\eta},\Vc{\xi}), \Vc{y}_{q}(\Vc{\eta}^{*},\Vc{\xi}^{*})-\Vc{y}_{eq}(\Vc{\eta}^{*},\Vc{\xi}^{*})\right>=0,
\end{align}
following from $\U \C =0$.
Without using the coordinate representation, this relation means that
\begin{align}
    \left<\Vc{x}_{p}-\Vc{x}_{eq}, \Vc{y}_{q}-\Vc{y}_{eq}\right>=0,
\end{align}
if $\Vc{x}_{p} \in \Polytope^{\X}(\Vc{\eta})$, $\Vc{x}_{q}^{\X}\in \Variety_{eq}^{\X}(\Vc{\xi}^{*})$, and $\Vc{x}_{eq}\in \Polytope^{\X}(\Vc{\eta}) \cap \Variety_{eq}^{\X}(\Vc{\xi}^{*})$.
For any $\Vc{x}, \Vc{x}'$ and $\Vc{x}''$, the divergence $\KL_{\X}[\Vc{x}\|\Vc{x}']$ satisfies  
\begin{align}
    \KL_{\X}[\Vc{x}\|\Vc{x}'] + &\KL_{\X}[\Vc{x}'\|\Vc{x}'']\notag \\
    &=\KL_{\X}[\Vc{x}\|\Vc{x}'']
    +\left<(\Vc{x}-\Vc{x}'),(\Vc{y}'-\Vc{y}'')\right>.
\end{align}
Thus, for $\Vc{x}_{p}$, $\Vc{x}_{q}$, and $\Vc{x}_{eq}$ satisfying $\Vc{x}_{p} \in \Polytope^{\X}(\Vc{\eta})$, $\Vc{x}_{q}^{\X}\in \Variety_{eq}^{\X}(\Vc{\xi}^{*})$, and $\Vc{x}_{eq}\in \Polytope^{\X}(\Vc{\eta}) \cap \Variety_{eq}^{\X}(\Vc{\xi}^{*})$,
the generalized Pythagorean theorem on $\X$ space holds (\fgref{fg:orthogonality} a):
\begin{align}
    \KL_{\X}[\Vc{x}_{p}\|\Vc{x}_{eq}] + \KL_{\X}[\Vc{x}_{eq}\|\Vc{x}_{q}] = \KL_{\X}[\Vc{x}_{p}\|\Vc{x}_{q}]. \label{eq:PythagoreanX}
\end{align}
This relation is geometric in the sense that it is independent of the choice of $\Vc{\eta}$ and $\Vc{\xi}^{*}$ or the choice of coordinate systems on $\Polytope^{\X}$ and $\Variety_{eq}$.

From the Pythagorean theorem, we obtain two variational characterizations of the equilibrium state (\fgref{fg:orthogonality}a ):
\begin{align}
    \Vc{x}_{eq}(\Vc{\eta},\Vc{\xi}^{*})&=\arg \min_{\Vc{x}\in \Polytope^{\X}(\Vc{\eta})} \KL_{\X}[\Vc{x}\|\Vc{x}'] \notag \\
    &\qquad \qquad \mbox{for any fixed $\Vc{x}' \in \Variety_{eq}^{\X}(\Vc{\xi}^{*})$} \label{eq:variational_x_1}\\
    &=\arg \min_{\Vc{x}'\in  \Variety_{eq}^{\X}(\Vc{\xi}^{*})} \KL_{\X}[\Vc{x}\|\Vc{x}'] \notag\\
    &\qquad \qquad \mbox{for any fixed $\Vc{x} \in \Polytope^{\X}(\Vc{\eta})$}. \label{eq:variational_x_2}
\end{align}
The former means that the equilibrium point can be obtained as the point in the stoichiometric polytope $\Polytope^{\X}$ at which $\KL_{\X}$ is minimized.
This is related to the relaxation process of $\Vc{x}(t)$ towards the equilibrium point in the stoichiometric compatibility class determined by the initial state $\Vc{x}_{0}$.

In contrast, the latter means that the equilibrium point, to which a given initial state $\Vc{x}_{0}$ converges, can be obtained as the point in the equilibrium variety $\Variety_{eq}^{\X}$ at which $\KL_{\X}$ is minimized.
The role of this equation will be clarified when we consider its thermodynamic meaning in \secref{sec:TD}.

Note that the convergence of $\Vc{x}(t)$ to $\Vc{x}_{eq}$ is not ensured by this geometric argument alone.
When mass action kinetic is assumed, one can directly prove the convergence by showing that the Bregman divergence $\KL_{\X}[\Vc{x}(t)\|\Vc{x}_{eq}]$ is a Lyapunov function \cite{shear1967JournalofTheoreticalBiology,higgins1968JournalofTheoreticalBiology,rao2016Phys.Rev.X}\footnote{We omit the proof because this result is commonly known.}:
\begin{align}
\frac{\dd \KL_{\X}[\Vc{x}(t)\|\Vc{x}_{eq}]}{\dd t}=- \Vc{j}(\Vc{x}(t))^{\Transpose}\ln\frac{\Vc{j}^{+}(\Vc{x}(t))}{\Vc{j}^{-}(\Vc{x}(t))}\le 0. \label{eq:Lyapunov}
\end{align}

\subsection{Variety, polytope, and orthogonality in the dual space}
Because $\X$ and $\Y$ are in bijection to each other, all the results obtained previously on $\X$ space can be carried over to $\Y$ space.
On $\Y$, the geometric nature of the stoichiometric polytope and the equilibrium \variety is swapped due to the logarithmic nature of the Legendre transform.
After transforming $\Polytope^{\X}(\Vc{\eta})$ and $\Variety^{\X}_{eq}(\Vc{\xi}^{*})$ by $\partial \varphi(\Vc{x})$, we obtain the stoichiometric \variety and the equilibrium polytope in $\Y$ space as (\fgref{fg:space_X} b and \fgref{fg:space_Y} a)
\begin{align}
\Variety^{\Y}(\Vc{\eta})&:=\partial \varphi\left(\Polytope^{\X}(\Vc{\eta})\right)\\
&=\{\Vc{y}|\Vc{y}=\ln\left[(\U^{*})^{\Transpose}\Vc{\eta} + \C \Vc{\xi}\right]-\hat{\Vc{y}},(\Vc{\xi},\Vc{\eta}) \in \mathcal{E} \}\notag\\
\Polytope^{\Y}_{eq}(\Vc{\xi}^{*})&:=\partial \varphi\left(\Variety^{\X}_{eq}(\Vc{\xi}^{*})\right)\label{eq:PY}\\
&=\left\{\Vc{y}|\Vc{y}=\C^{*}\Vc{\xi}^{*} + \U^{\Transpose}\Vc{\eta}^{*},\Vc{\eta}^{*}\in\mathbb{R}^{\ell}\right\}.\notag
\end{align}

Now, the curved \variety $\Variety^{\X}_{eq}(\Vc{\xi}^{*})$ in $\X$ space is a flat polytope $\Polytope^{\Y}_{eq}(\Vc{\xi}^{*})$ in $\Y$ space.
Conversely, the flat polytope $\Polytope^{\X}(\Vc{\eta})$ in $\X$ is a curved manifold $\Variety^{\Y}(\Vc{\eta})$ in $\Y$\footnote{The space $\Variety^{\Y}(\Vc{\eta})$ is not necessarily an algebraic variety because it is not defined by algebraic equations. Thus, we use the word manifold here. }.
This is the essence of the dually flat structure.
In the conventional formulation of information geometry by Amari \cite{amari2016}, the theory is formulated by using only $\X$ space, which obscures the relation between $\X$ and $\Y$.
Moreover, as we will see in \secref{sec:TD}, the relation becomes fundamental when we consider thermodynamics.

In $\Y$ space, for $\Vc{y}_{p}$, $\Vc{y}_{q}$, and $\Vc{y}_{eq}$ satisfying $\Vc{y}_{p} \in \Variety^{\Y}(\Vc{\eta})$, $\Vc{y}_{q}^{\Y}\in \Polytope_{eq}^{\Y}(\Vc{\xi}^{*})$, and $\Vc{y}_{eq}\in \Variety^{\Y}(\Vc{\eta}) \cap \Polytope_{eq}^{\Y}(\Vc{\xi}^{*})$, the Pythagorean relation and the corresponding variational relations (\fgref{fg:orthogonality} b) are given by
\begin{align}
    \KL_{\Y}[\Vc{y}_{q}\|\Vc{y}_{eq}] + \KL_{\Y}[\Vc{y}_{eq}\|\Vc{y}_{p}] = \KL_{\Y}[\Vc{y}_{q}\|\Vc{y}_{p}]. \label{eq:PythagoreanY}
\end{align}
Analogously to \eqnref{eq:variational_x_1} and \eqnref{eq:variational_x_2}, we have two variational characterizations of the equilibrium state in $\Y$ space (\fgref{fg:orthogonality} b) as 
\begin{align}
    \Vc{y}_{eq}(\Vc{\xi}^{*},\Vc{\eta})&=\arg \min_{\Vc{y}\in \Variety^{\Y}(\Vc{\xi}^{*})} \KL_{\Y}[\Vc{y}'\|\Vc{y}] \notag \\
    &\qquad \qquad \mbox{for any $\Vc{y}' \in \Polytope_{eq}^{\Y}(\Vc{\eta})$},\label{eq:variational_y_1} \\
    &=\arg \min_{\Vc{y}'\in  \Polytope_{eq}^{\Y}(\Vc{\eta})} \KL_{\Y}[\Vc{y}'\|\Vc{y}] \notag \\
    &\qquad \qquad \mbox{for any $\Vc{y} \in \Variety^{\Y}(\Vc{\xi}^{*})$}. \label{eq:variational_y_2}
\end{align}
The implication of these equation, especially that of the latter, will be discussed in \secref{sec:TD} after clarifying their connection to thermodynamics.

\section{Differential Geometric Structure}\label{sec:DG}
In the previous section, we have obtained the geometric structure by exploiting only the information of the first derivative of $\varphi(\Vc{x})$. 
In this section, we explore the information contained in the second derivative of $\varphi(\Vc{x})$, i.e., its Hessian.

\subsection{Hessian and Fisher information}
Because $\varphi(\Vc{x})$ is strongly convex, its Hessian (second derivative matrix)
\begin{align}
    \FM_{\X}(\Vc{x})\defeq \left(\frac{\partial^{2}\varphi(\Vc{x})}{\partial x_{i}\partial x_{j}}\right)_{i,j}=\diag \left(\frac{1}{\Vc{x}}\right),
\end{align}
is positive definite, which endows the tangent space of $\X$ with a natural inner product structure:
\begin{align}
    \left<\Delta\Vc{x},\Delta\Vc{x}'\right>_{\Vc{x}} \defeq \left<\FM_{\X}(\Vc{x})\Delta\Vc{x},\Delta \Vc{x}' \right>,\label{eq:innerproduct_X}
\end{align}
where $\Delta\Vc{x}$ and $\Delta\Vc{x}'$ are elements of the tangent space $\Tan_{\Vc{x}} \X$ at $\Vc{x}\in\X$.
Moreover, $\left<\cdot,\cdot\right>$ is the bilinear form defined on $\Tan_{\Vc{x}} \X \times \Tan_{\Vc{x}}^{*} \X$, where $\coTan_{\Vc{x}} \X$ is the corresponding cotangent space.
The induced norm is $\left|\Delta\Vc{x}\right|_{\Vc{x}} \defeq \sqrt{\left<\Delta\Vc{x},\Delta\Vc{x}\right>_{\Vc{x}}}$.
Also, $\FM_{\X}(\Vc{x})$ induces a mapping from the tangent space $\Tan_{\Vc{x}} \X$ to the cotangent space $\coTan_{\Vc{x}} \X$ via $\Delta \Vc{y}=\FM_{\X}(\Vc{x})\Delta\Vc{x} \in \coTan_{\Vc{x}} \X$.

The Hessian of $\varphi^{*}(\Vc{y})$ is computed as
\begin{align}
    \FM_{\Y}(\Vc{y})\defeq \left(\frac{\partial^{2}\varphi^{*}(\Vc{y})}{\partial y_{i}\partial y_{j}}\right)_{i,j}=\diag\,  e^{\Vc{y}+\hat{\Vc{y}}}
\end{align}
and satisfies $\FM_{\X}(\Vc{x})\FM_{\Y}(\Vc{y})=\identityM$ because $\Vc{x}=e^{\Vc{y}+\hat{\Vc{y}}}$. 
Thus, $\FM_{\X}^{-1}(\Vc{x})=\FM_{\Y}(\Vc{y})$ and $\FM_{\Y}^{-1}(\Vc{y})=\FM_{\X}(\Vc{x})$.
The inner product induced by $\FM_{\Y}(\Vc{y})$ on the dual tangent space $\Tan_{\Vc{y}} \Y$ is 
\begin{align}
    \left<\Delta\Vc{y},\Delta\Vc{y}'\right>_{\Vc{y}} \defeq \left<\FM_{\Y}(\Vc{y})\Delta\Vc{y},\Delta \Vc{y}' \right>.\label{eq:innerproduct_Y}
\end{align}
Because of the linear duality between $\X$ and $\Y$, we have relations: $\Tan_{\Vc{x}}\cong \coTan_{\Vc{y}}$ and $\coTan_{\Vc{x}}\cong \Tan_{\Vc{y}}$.

In information geometry, $\FM_{\X}(\Vc{x})$ is known as Fisher information.
It is related to the infinitesimal change of the Kullback-Leibler divergence as
\begin{align}
    \dd s^{2}&=2 \KL_{\X}[\Vc{x}\|\Vc{x}+\Delta \Vc{x}] & =\left<\Delta \Vc{x},\FM_{\X}(\Vc{x}) \Delta \Vc{x}\right>\\
    &=2 \KL_{\Y}[\Vc{y} + \Delta \Vc{y}\|\Vc{y}] & =\left<\Delta \Vc{y},\FM_{\Y}(\Vc{y}) \Delta \Vc{y}\right>.
\end{align}

\subsection{Fisher information for the dual and mixed coordinates}

By inserting $\Delta \Vc{x} = (\U^{*})^{\Transpose} \Delta \Vc{\eta} + \C \Delta \Vc{\xi}$ or $\Delta \Vc{y}= \U^{\Transpose} \Delta \Vc{\eta}^{*} + \C^{*} \Delta \Vc{\xi}^{*} $, 
we obtain Fisher information matrices for $(\Vc{\eta},\Vc{\xi})$ or $(\Vc{\eta}^{*},\Vc{\xi}^{*})$:
\begin{align}
    \FM_{\Vc{\eta},\Vc{\xi}}(\Vc{x}) &\defeq \begin{pmatrix}
    \U^{*}\FM_{\X}(\Vc{x})(\U^{*})^{\Transpose} & \U^{*}\FM_{\X}(\Vc{x})\C\\
    \C^{\Transpose}\,\FM_{\X}(\Vc{x})(\U^{*})^{\Transpose} & \C^{\Transpose}\FM_{\X}(\Vc{x})\C \label{eq:FM_X}
\end{pmatrix}
\end{align}
\begin{align}
    \FM_{\Vc{\eta}^{*},\Vc{\xi}^{*}}(\Vc{y}) &\defeq \begin{pmatrix}
    \U\FM_{\Y}(\Vc{y})\U^{\Transpose} & \U\FM_{\Y}(\Vc{y})\C^{*}\\
    (\C^{*})^{\Transpose}\,\FM_{\Y}(\Vc{y})\U^{\Transpose} & (\C^{*})^{\Transpose}\FM_{\Y}(\Vc{y})\C^{*} \label{eq:FM_Y}
\end{pmatrix}
\end{align}
We can verify that $\FM_{\Vc{\eta}^{*},\Vc{\xi}^{*}}(\Vc{y})$ is the inverse matrix of $\FM_{\Vc{\eta},\Vc{\xi}}(\Vc{x})$ by directly computing $\FM_{\Vc{\eta}^{*},\Vc{\xi}^{*}}(\Vc{y})\FM_{\Vc{\eta},\Vc{\xi}}(\Vc{x})=\identityM$  where we use the fact that
\begin{align}
    \ProjM_{\U}\defeq \U^{\Transpose}\U^{*},\qquad \ProjM_{\C}\defeq\C^{*}\C^{\Transpose}
\end{align}
are orthogonal projection matrices and satisfy $\ProjM_{\U} + \ProjM_{\C}^{\Transpose} = \identityM$.
The effective metric matrix can be further simplified by using mixed coordinates.
In this case $\Delta \Vc{x}$ is represented by
\begin{align}
    \Delta \Vc{x} &= (\U^{*})^{\Transpose} \Delta \Vc{\eta} + \FM_{\Y}(\boldsymbol{y})\C^{*} \Delta \Vc{\xi}^{*}\\
    &= \FM_{\Y}(\Vc{x})\U^{\Transpose} \Delta \Vc{\eta}^{*} + \C \Delta \Vc{\xi}.
\end{align}
The cross terms in $\dd s^{2}$ disappear due to the dual orthogolality:
\begin{align}
    \dd s^{2} \sim & \left<\Delta \Vc{\eta}, \FM_{\Vc{\eta}}(\Vc{x}) \Delta \Vc{\eta}\right> + \left<\Delta \Vc{\xi}^{*}, \FM_{\Vc{\xi}^{*}}(\Vc{y}) \Delta \Vc{\xi}^{*}\right>,\\ 
    \sim & \left<\Delta \Vc{\eta}^{*}, \FM_{\Vc{\eta}^{*}}(\Vc{y}) \Delta \Vc{\eta}^{*}\right> + \left<\Delta \Vc{\xi}, \FM_{\Vc{\xi}}(\Vc{x}) \Delta \Vc{\xi}\right>,
\end{align}
where $\FM_{\Vc{\eta}}(\Vc{x}) \defeq \U^{*}\FM_{\X}(\Vc{x})(\U^{*})^{\Transpose}$ and $\FM_{\Vc{\xi}}(\Vc{x})\defeq \C^{\Transpose}\FM_{\X}(\Vc{x})\C$ are the diagonal blocks of $\FM_{\Vc{\eta},\Vc{\xi}}(\Vc{x})$ (\eqnref{eq:FM_X}), whereas $\FM_{\Vc{\eta}^{*}}(\Vc{y})\defeq \U\FM_{\Y}(\Vc{y})\U^{\Transpose}$ and $\FM_{\Vc{\xi}^{*}}(\Vc{y}) \defeq (\C^{*})^{\Transpose}\FM_{\Y}(\Vc{y})\C^{*}$ are those of $\FM_{\Vc{\eta}^{*},\Vc{\xi}^{*}}(\Vc{y})$ (\eqnref{eq:FM_Y}).
This is a tangent and cotangent space version of the generalized Pythagorean relation and the orthogolonality between the stoichiometric polytopes and equilibrium varieties.
\footnote{Note that neither $\FM_{\Vc{\eta}}(\Vc{x}) \FM_{\Vc{\eta}^{*}}(\Vc{y}) = \identityM$ nor $\FM_{\Vc{\xi}}(\Vc{x})\FM_{\Vc{\xi}^{*}}(\Vc{y}) = \identityM$ holds.}

\section{Link to chemical thermodynamics}\label{sec:TD}
In this section, we clarify how the Hessian geometric structure and its constituents can be related to equilibrium chemical thermodynamics. 
To this end, because the results were derived kinetically from the law of mass action and detailed balancing, we have to rederive the same results from thermodynamics, if at all possible, without assuming any kinetics. 
This is achieved in our accompanying paper.
Here, we make the correspondence precise. 
Refer to \cite{sughiyama2021ArXiv211212403Cond-MatPhysicsphysics} for the more general results derived from thermodynamics. 

\subsection{Chemical potential and Free energy}
The chemical potential of a dilute solution (or equivalently of an ideal gas) is given by
\begin{align}
    \Vc{\mu}(\Vc{x})=\hat{\Vc{\nu}}^{o}(\T)+\R \T \ln \Vc{x},
\end{align}
where $\T$ is the temperature of the system, $\R$ is the gas constant, and $\hat{\Vc{\nu}}^{o}(\T)$ is the standard chemical potential of $\Vc{x}$\footnote{We used $\hat{\Vc{\nu}}^{o}(\T)$ for the standard chemical potential rather than more conventional $\hat{\Vc{\mu}}^{o}(\T)$ to make the notation consistent with those in \cite{sughiyama2021ArXiv211212403Cond-MatPhysicsphysics}}.
The vector $\Vc{y}=\ln \Vc{x} - \hat{\Vc{y}}$ is related to the chemical potential via
\begin{align}
    \Vc{y}=\Vc{\mu}/(\R \T), \qquad \hat{\Vc{y}} = - \hat{\Vc{\nu}}^{o}/(\R \T),
\end{align}
which corresponds to \com{equation (63) in \cite{sughiyama2021ArXiv211212403Cond-MatPhysicsphysics}.}
Thus, $\Y$ is the space of chemical potentials, which is thermodynamically conjugate to the concentration space of molecular species $\X$.
The potential function $\varphi(\Vc{x})$ is associated with the Gibbs free energy of the system (without reservoir):
\begin{align}
\R\T\varphi(\Vc{x})&= \R\T\left[\ln\Vc{x}-\hat{\Vc{y}}-\Vc{1} \right]^{\Transpose}\Vc{x}\\
&=\left[\Vc{\mu}(\Vc{x})-\R \T \Vc{1} \right]^{\Transpose}\Vc{x}=\Gibbs_{sys}(\boldsymbol{x})-\Gibbs_{0},
\end{align}
where $\Gibbs_{0}$ is constant. This equation corresponds to \com{equation (59) in \cite{sughiyama2021ArXiv211212403Cond-MatPhysicsphysics}} under the additional assumption that there are no molecules exchanged with the reservoir. Thus, the system is closed.
Because $\Gibbs_{sys}(\Vc{x})$ is minimized for $\ln\hat{\Vc{x}}=\hat{\Vc{y}}$, which satisfies $\Vc{\mu}(\hat{\Vc{x}})=0$, such that $\hat{\Vc{\nu}}^{o}$, or equivalently $\hat{\Vc{y}}$, specifies the intrinsic equilibrium state that is attained if the system is closed and free from stoichiometric constraints.
We used the term Gibbs free energy for $\Gibbs_{sys}(\boldsymbol{x})$, following the convention of chemical thermodynamics\footnote{
More precisely, $\Gibbs_{sys}(\boldsymbol{x})$ is a Helmholtz free energy if the volume is predominantly determined by the non-reactive solvent as is assumed in the theory of chemical reaction systems  (cf. \com{Discussion in \cite{sughiyama2021ArXiv211212403Cond-MatPhysicsphysics}}).}.

\subsection{Total entropy and Bregman divergence}
Next, consider the case that the system is open but free from stoichiometric constraints, i.e., $\Ker[\stoiMatrix^{\Transpose}]=\{0\}$.
The total entropy of the system is related to $\varphi(\Vc{x})$ by
\begin{align}
    \Sigma^{tot}(\Vc{x};\tilde{\Vc{y}}_{res})= \R \Volume \left[\left<\tilde{\Vc{y}}_{res},\Vc{x}\right>-\varphi(\Vc{x}) \right] + \mathrm{const}, \label{eq:tot_entropy}
\end{align}
where $\tilde{\Vc{y}}_{res}$ is uniquely specified by the state of the chemical reservoir (\com{see equation (41) in \cite{sughiyama2021ArXiv211212403Cond-MatPhysicsphysics}}) through the equation
\begin{align}
    \tilde{\Vc{\mu}}^{\Transpose} O = -\tilde{\Vc{y}}_{res}^{\Transpose}\stoiMatrix.\label{eq:y_p_cond}
\end{align}
Here, $\tilde{\Vc{\mu}}$ is the chemical potential of molecules, which can be exchanged with reservoir and $O$ is the stoichiometric matrix for these exchange reactions.
The existence of $\tilde{\Vc{y}}_{res}$ follows from the equilibrium parameter condition (\eqnref{eq:K}) and \com{Eq (78) of \cite{sughiyama2021ArXiv211212403Cond-MatPhysicsphysics}.}
The uniqueness of $\tilde{\Vc{y}}_{res}$ is assured by $\Ker[\stoiMatrix^{\Transpose}]=\{0\}$.
If $\Vc{x}=\tilde{\Vc{x}}_{res}$, we have
\begin{align}
    \Sigma^{tot}(\tilde{\Vc{x}}_{res};\tilde{\Vc{y}}_{res})= \R \Volume \varphi^{*}(\tilde{\Vc{y}}_{res}) + \mathrm{const},
\end{align}
which indicates that $\varphi^{*}(\Vc{y})$ is the total entropy at the equilibrium point specified by the parameter $\tilde{\Vc{y}}_{res}$.
By using the relation between the potential $\varphi(\Vc{x})$ and the Bregman divergence $\KL_{\X,\Y}[\Vc{x}\|\Vc{y}']$, the difference of total entropy between $\Vc{x}$ and $\Vc{x}'$ can be associated with the difference of Bregman divergences
\begin{align}
\Sigma^{tot}(\Vc{x};\tilde{\Vc{y}}_{res})& -\Sigma^{tot}(\Vc{x}';\tilde{\Vc{y}}_{res})\notag \\
&= -\R \Volume\left[\KL_{\X,\Y}[\Vc{x}\|\tilde{\Vc{y}}_{res}]-\KL_{\X,\Y}[\Vc{x}'\|\tilde{\Vc{y}}_{res}]\right].\label{eq:Entropy_dif}
\end{align}
Without stoichiometric constraints, $\Sigma^{tot}(\Vc{x};\tilde{\Vc{y}}_{res})$ is maximized at the corresponding equilibrium state $\Vc{x}_{eq}$
\begin{align}
    \tilde{\Vc{x}}_{eq} \defeq \arg \max_{\Vc{x}}\Sigma^{tot}(\Vc{x};\tilde{\Vc{y}}_{res}).
\end{align}
By inserting $\Vc{x}'=\tilde{\Vc{x}}_{res}$ into \eqnref{eq:Entropy_dif}, where $\tilde{\Vc{x}}_{res}$ is the Legendre transform of $\tilde{\Vc{y}}_{res}$, we obtain
\begin{align}
\Sigma^{tot}(\Vc{x};\tilde{\Vc{y}}_{res})& = \Sigma^{tot}(\tilde{\Vc{x}}_{res};\tilde{\Vc{y}}_{res}) -\R \Volume\KL_{\X,\Y}[\Vc{x}\|\tilde{\Vc{y}}_{res}].
\end{align}
Thus
\begin{align}
    \Vc{x}_{eq} =\arg \min_{\Vc{x}}\KL_{\X,\Y}[\Vc{x}\|\tilde{\Vc{y}}_{res}] = \tilde{\Vc{x}}_{res}\label{eq:eq_no_constraint}
\end{align}
and the total entropy production is 
\begin{align}
    \Delta \Sigma^{tot}(\Vc{x}\to\Vc{x}_{eq};\tilde{\Vc{y}}_{res})&\defeq\Sigma^{tot}(\tilde{\Vc{x}}_{res};\tilde{\Vc{y}}_{res})-\Sigma^{tot}(\Vc{x};\tilde{\Vc{y}}_{res})\notag \\
    &= \R \Volume\KL_{\X,\Y}[\Vc{x}\|\tilde{\Vc{y}}_{res}].
\end{align}
Equation (\ref{eq:eq_no_constraint}) indicates that specifying a certain equilibrium parameter $\tilde{\Vc{y}}\in \Y$ by modulating the reservoir is equivalent to specifying the state $\tilde{\Vc{x}} \in \X$ that is the equilibrium state of the system under the reservoir parameter $\tilde{\Vc{y}}$ because of the one-to-one correspondence between $\tilde{\Vc{x}}$ and $\tilde{\Vc{y}}$.

\subsection{Entropy production under constraints}
If stoichiometric constraints exist, i.e., $\Ker[\stoiMatrix^{\Transpose}]\neq\{0\}$, then $\tilde{\Vc{y}}_{res}$ cannot be determined uniquely from the reservoir parameter by \eqnref{eq:y_p_cond}.
Nevertheless, \eqnref{eq:tot_entropy} has meaning.
Let $\tilde{\Vc{y}}$ be a particular solution satisfying \eqnref{eq:y_p_cond}.
A system $\Vc{x}(t)$ with initial condition $\Vc{x}_{0}$ is restricted to the stoichiometric polytope $\Polytope^{\X}(\Vc{\eta}_{0})$ which contains $\Vc{x}_{0}$.
Then, the equilibrium state is characterized as 
\begin{align}
    \Vc{x}_{eq}=&\arg \max_{\Vc{x}\in \Polytope_{\X}(\Vc{\eta}_{0})}\Sigma^{tot}(\Vc{x};\tilde{\Vc{y}})\\
    =&\arg \min_{\Vc{x}\in \Polytope_{\X}(\Vc{\eta}_{0})}\KL_{\X,\Y}[\Vc{x}\|\tilde{\Vc{y}}].
\end{align}
Let $\Variety^{\X}_{eq}(\tilde{\Vc{\xi}}^{*})$ be the equilibrium \variety to which $\tilde{\Vc{y}}$ belongs, i.e., $\tilde{\Vc{x}}$, the Legendre transform of $\tilde{\Vc{y}}$, satisfies $\tilde{\Vc{x}} \in \Variety^{\X}_{eq}(\tilde{\Vc{\xi}}^{*})$.
Moreover, let $\Vc{x}_{int}$ be the intersection point of the polytope and the \variety, i.e., $\Vc{x}_{int}\in \Polytope_{\X}(\Vc{\eta}_{0})\cap\Variety^{\X}_{eq}(\tilde{\Vc{\xi}}^{*})$. 
As shown before, this intersection is unique and can be specified as $\Vc{x}_{int}=\Vc{x}_{int}(\Vc{\eta}_{0},\tilde{\Vc{\xi}}^{*})$.
For any $\Vc{x} \in \Polytope_{\X}(\Vc{\eta}_{0})$, the generalized Pythagorean relation holds: 
\begin{align}
\KL_{\X,\Y}[\Vc{x}\|\tilde{\Vc{y}}]= \KL_{\X,\Y}[\Vc{x}\|\Vc{y}_{int}] + \KL_{\X,\Y}[\Vc{x}_{int}\|\tilde{\Vc{y}}].
\end{align}
Thus, 
\begin{align}
    \tilde{\Vc{x}}_{eq}=&\arg \min_{\Vc{x}\in \Polytope_{\X}(\Vc{\eta}_{0})}\left[\KL_{\X,\Y}[\Vc{x}\|\Vc{y}_{int}] + \KL_{\X,\Y}[\Vc{x}_{int}\|\tilde{\Vc{y}}] \right]\notag \\
    =&\arg \min_{\Vc{x}\in \Polytope_{\X}(\Vc{\eta}_{0})}\KL_{\X,\Y}[\Vc{x}\|\Vc{y}_{int}]=\Vc{x}_{int}.
\end{align}
This verifies that the equilibrium state $\Vc{x}_{eq}$, which maximizes the total entropy, is characterized by the intersection point of the polytope and variety.
Also, the entropy production becomes 
\begin{align}
   \Delta \Sigma^{tot}(\Vc{x}\to\tilde{\Vc{x}}_{eq};\tilde{\Vc{y}})&\defeq\Sigma^{tot}(\Vc{x}_{eq};\tilde{\Vc{y}})-\Sigma^{tot}(\Vc{x};\tilde{\Vc{y}})\notag \\
    &= \R \Volume\KL_{\X,\Y}[\Vc{x}\|\Vc{y}_{eq}],
\end{align}
where the relevant part is independent of $\tilde{\Vc{y}}$ owning to the Pythagorean relation. 
The convergence to $\Vc{x}_{eq}$ is then attributed to the second law, which was also provens kinetically for mass action system (\eqnref{eq:Lyapunov}).
The independence of the choice of $\tilde{\Vc{y}}$ can be understood more clearly in $\Y$ space from the dual variational equation:
\begin{align}
    \Vc{y}_{eq}=&\underset{\tilde{\Vc{y}}\in \Polytope^{\Y}_{eq}(\tilde{\Vc{\xi}}^{*}) }{\arg\min}\left[\KL_{\X,\Y}[\Vc{x}\|\Vc{y}_{int}] + \KL_{\X,\Y}[\Vc{x}_{int}\|\tilde{\Vc{y}}] \right]\notag \\
    &=\underset{\tilde{\Vc{y}}\in \Polytope^{\Y}_{eq}(\tilde{\Vc{\xi}}^{*}) }{\arg\min}\KL_{\X,\Y}[\Vc{x}_{int}\|\tilde{\Vc{y}}]=\Vc{y}_{int},
\end{align}
which corresponds to \eqnref{eq:variational_y_2}.
From the definition of $\Polytope^{\Y}_{eq}(\tilde{\Vc{\xi}}^{*})$ by \eqnref{eq:PY}, any $\tilde{\Vc{y}},\tilde{\Vc{y}}' \in \Polytope^{\Y}_{eq}(\tilde{\Vc{\xi}}^{*})$ satisfies  $(\tilde{\Vc{y}}-\tilde{\Vc{y}}')^{\Transpose}\stoiMatrix=0$.
Thus, the choice of a particular $\tilde{\Vc{y}}$ can only contribute to  $\KL_{\X,\Y}[\Vc{x}_{int}\|\tilde{\Vc{y}}]$, which is orthogonal to $\KL_{\X,\Y}[\Vc{x}\|\Vc{y}_{int}]$.
In other words, if there exist stoichiometric constraints, the relevant quantity specified by the reservoir is no longer a point or value but a geometric object, i.e., the whole equilibrium \variety.

\subsection{Linear response of total entropy}
Finally, we investigate responses of the total entropy to infinitesimal changes of either $\Vc{x}$ or $\Vc{y}_{res}$.
First, suppose that state $\Vc{x}$ is perturbed to $\Vc{x}'=\Vc{x}+\Delta \Vc{x}$. For a general perturbation, which is not restricted by stoichiometric constraints, we obtain
\begin{align}
    \frac{\Delta \Sigma^{tot}(\Vc{x}\to\Vc{x}';\tilde{\Vc{y}}_{res})}{\R\Volume}&=\KL_{\X}[\Vc{x}+\Delta \Vc{x}\|\Vc{x}_{eq}]-\KL_{\X}[\Vc{x}\|\Vc{x}_{eq}]\notag \\
    &\approx\left<\Delta \Vc{x},\Vc{y}-\Vc{y}_{eq} \right>. \label{eq:Ent_perturb_X}
\end{align}
Here, $\Delta \Vc{x}$ and $\Vc{y}-\Vc{y}_{eq}$ are treated as elements of $\X$ and $\Y$, respectively.
Moreover, we use an implicit identification of the base spaces with their tangent and cotangent spaces.
Because $\X$ and $\Y$ inherit the affine structure of dual vector spaces $\mathbb{V}$ and $\mathbb{V}^{*}$, and because the tangent and cotangent spaces are isomorphic to these vector spaces, and we can identify these spaces (non-canonically) as
\begin{align}
 \coTan_{\Vc{y}} \Y \cong \Tan_{\Vc{x}} \X \cong \mathbb{V} \supset \X ,\quad \quad \coTan_{\Vc{x}} \X \cong \Tan_{\Vc{y}}\Y  \cong \mathbb{V}^{*} = \Y. \notag
\end{align}
Because of this isomorphism, we can also regard $\Delta \Vc{x}$ and $\Vc{y}-\Vc{y}_{eq}$ as elements of tangent and cotangent spaces.
By combining this fact with the Cauchy-Schwartz inequality, we obtain
\begin{align}
    \left|\frac{\Delta \Sigma^{tot}(\Vc{x}\to\Vc{x}';\tilde{\Vc{y}}_{res})}{\R\Volume}\right|&=\left|\left<\Delta \Vc{x}, \Vc{y}-\Vc{y}_{eq} \right>\right|\\
    &\le \left|\Vc{y}-\Vc{y}_{eq}\right|_{\Vc{y}}\left|\Delta \Vc{x}\right|_{\Vc{x}},
\end{align}
where $\left|\cdot\right|_{\Vc{x}}$ and $\left|\cdot\right|_{\Vc{y}}$ are the metrics on the tangent and cotangent spaces induced by the Fisher information (Eqs. \ref{eq:innerproduct_X} and \ref{eq:innerproduct_Y}). 
From this, we see that $\left|\Vc{y}-\Vc{y}_{eq}\right|_{\Vc{y}}$ is an upper bound of the sensitivity of entropy production proposed in \cite{yoshimura2021Phys.Rev.Research} and that entropy production is maximized if $\Delta \Vc{x}$ is parallel to $\Vc{y}-\Vc{y}_{eq}$ in the sense of linear duality.

\section{Complex-balanced systems}\label{sec:CB}
The Hessian geometric structure can also be extended beyond equilibrium chemical systems to complex balanced systems.

Complex balanced systems, introduced by Horn and Jackson \cite{horn1972Arch.RationalMech.Anal.a}, are a class of nonequilibrium systems, which preserve several properties of equilibrium chemical systems. 
A complex balanced system has a unique steady state called complex balanced state, which is also globally stable \cite{gopalkrishnan2014SIAMJ.Appl.Dyn.Syst.,craciun2016ArXiv150102860Math}.
Moreover, the generalized Kullback-Leibler divergence works as the Lyapunov function of the system \cite{horn1972Arch.RationalMech.Anal.a}.
This similarity is partially attributed to the shared geometric structure between equilibrium and complex balanced systems.

To define the complex balanced state, we note that the stoichiometric matrix $\stoiMatrix$ can be decomposed as $\stoiMatrix=-Y \IncMatrix$.
Hereby, $\IncMatrix$ is the incidence matrix of the oriented graph obtained by regarding the sets of reactants or products in the reaction network as vertices (also called complexes) and the reactions as edges.
The orientation of an edge is determined by the direction of the corresponding forward reaction.
$Y$ maps the complexes to the respective constituent molecular species.
We also assume that the reaction flux $\Vc{\flux}^{\pm}(\Vc{x}_{cb};\Vc{\theta})$ satisfies the law of mass action.

Then the set of complex balanced states is given by
\begin{align}
\Variety_{cb}^{\X}(\Vc{\theta})\defeq  \{\Vc{x}_{cb}> 0|\IncMatrix\Vc{\flux}(\Vc{x}_{cb};\Vc{\theta})=\Vc{0}\}.    
\end{align}
Because $\Vc{\flux}^{\pm}(\Vc{x}_{cb};\Vc{\theta})$ consists of monominals, $\Variety_{cb}^{\X}(\Vc{\theta})$ is an algebraic variety.
As before, we define the parameter sets $\Theta_{cb}$ in which $\Variety_{cb}^{\X}(\Vc{\theta})$ is non-empty, i.e.:
\begin{align}
\Theta_{cb} \defeq \{\Vc{\theta}| \Variety_{cb}^{\X}(\Vc{\theta})\neq \emptyset\}.
\end{align}
Obviously, $\Theta_{eq} \subset \Theta_{cb}$ and thus an equilibrium variety is a special class of complex balanced varieties.
Compared with the definition of an equilibrium variety, $\IncMatrix\Vc{\flux}(\Vc{x}_{cb};\Vc{\theta})=\Vc{0}$ is in general not given by binomial equations. 
Nevertheless, $\Variety_{cb}^{\X}(\Vc{\theta})$ is a toric variety, meaning that $\IncMatrix\Vc{\flux}(\Vc{x}_{cb};\Vc{\theta})=\Vc{0}$ can be converted to binomial equations by appropriate algebraic manipulation similar to the Gaussian elimination for linear equations \cite{craciun2009JournalofSymbolicComputation}.
In other words, a complex balanced state is defined by hidden detailed balance (binomial) equations.
Thus, as a toric variety, $\Variety_{cb}^{\X}(\Vc{\theta})$ has an explicit parametrization in the same way as the equilibrium variety (\eqnref{eq:VarEq}).
Moreover, the complex balanced variety has the same design matrix and thus is given by
\begin{align}
    \Variety_{cb}^{\X}(\Vc{\theta})&=\left\{\Vc{x}|\ln \Vc{x}=\ln \tilde{\Vc{x}}_{cb} + \U^{\Transpose}\Vc{\eta}^{*}\right\}, \label{eq:VarCb}
\end{align}
where $\tilde{\Vc{x}}_{cb}$ is determined by the actual values of kinetic parameters \cite{craciun2009JournalofSymbolicComputation}.
It should be noted that multiple parameter values $\Vc{\theta}, \Vc{\theta}' \in \Xi_{\theta}$ can generate the same variety $\Variety_{cb}^{\X}(\Vc{\theta}) = \Variety_{cb}^{\X}(\Vc{\theta}')$. 

As geometrical objects, the complex balanced variety and equilibrium variety are indistinguishable and therefore the embedding of each complex balanced variety into $\X$ is parametrized analogously by $\Vc{\xi}^{*}$. From now on, we write $\Variety_{cb}^{\X}(\Vc{\xi}^{*})$ for the complex balanced variety embedded in $\X$.
As a result, the Hessian geometric structure of the equilibrium variety described in \secref{sec:HG} is preserved in the more general case of complex balance.

For example, for a given initial state $\Vc{x}_{0} \in \Polytope^{\X}(\Vc{\eta})$, the complex balanced steady state $\Vc{x}_{cb}$ is the intersection of the polytope $\Polytope^{\X}(\Vc{\eta})$, determined by the initial state, and the complex balanced variety  $\Variety_{cb}^{\X}(\Vc{\xi}^{*})$:
\begin{align}
    \Vc{x}_{cb} \in \Polytope^{\X}(\Vc{\eta}) \cap \Variety_{cb}^{\X}(\Vc{\xi}^{*}). \label{eq:Birch_point_cb}
\end{align}
For the same reason as the equilibrium state, $\Vc{x}_{cb}$ is unique and the intersection is transversal \cite{craciun2009JournalofSymbolicComputation}.
By using the same potential function $\varphi(\Vc{x})$ as in \eqnref{eq:potential_function}, define the dual space $\Y$, the Legendre transformed function $\varphi^{*}(\Vc{x})$, the Bregman divergence $\KL_{\X}$, the tangent and cotangent spaces $\Tan\X$ and $\Tan\Y$, and also the Fisher information given by $\FM_{\X}(\Vc{x})$ and $\FM_{\Y}(\Vc{y})$.
In other words, we can naturally embed a $\Variety_{cb}^{\X}(\Vc{\theta})$ into the Hessian geometric structure constructed for the equilibrium varieties.
As a result, the same Pythagorean relation as \eqnref{eq:PythagoreanX} holds for the complex balanced case:
\begin{align}
    \KL_{\X}[\Vc{x}_{p}\|\Vc{x}_{cb}] + \KL_{\X}[\Vc{x}_{cb}\|\Vc{x}_{q}] = \KL_{\X}[\Vc{x}_{p}\|\Vc{x}_{q}], \label{eq:PythagoreanX_CB}
\end{align}
where $\Vc{x}_{p} \in \Polytope^{\X}(\Vc{\eta})$, $\Vc{x}_{q}^{\X}\in \Variety_{cb}^{\X}(\Vc{\xi}^{*})$, and $\Vc{x}_{cb}\in \Polytope^{\X}(\Vc{\eta}) \cap \Variety_{cb}^{\X}(\Vc{\xi}^{*})$.
Thus, the complex balanced state admits the variational characterizations
\begin{align}
    \Vc{x}_{cb}(\Vc{\eta},\Vc{\xi}^{*})&=\arg \min_{\Vc{x}\in \Polytope^{\X}(\Vc{\eta})} \KL_{\X}[\Vc{x}\|\Vc{x}'] \notag \\
    &\qquad \qquad \mbox{for any fixed $\Vc{x}' \in \Variety_{cb}^{\X}(\Vc{\xi}^{*})$}\\
    &=\arg \min_{\Vc{x}'\in  \Variety_{cb}^{\X}(\Vc{\xi}^{*})} \KL_{\X}[\Vc{x}\|\Vc{x}'] \notag\\
    &\qquad \qquad \mbox{for any fixed $\Vc{x} \in \Polytope^{\X}(\Vc{\eta})$}.
\end{align}
The analogous variational characterization also holds in the dual space.

While the geometric structure is inherited, the thermodynamic aspects are not. 
For example, the space $\Y$ may not admit the same interpretation as the chemical potential space. 
The potential function $\varphi(\Vc{x})$ and Bregman divergence $\KL_{\X}$ are no longer associated with the thermodynamic potential function and the difference of total entropy. 
All of this is rooted in the general lack of a thermodynamically consistent characterization of nonequilibrium states.
While there have been continuous attempts to achieve a thermodynamic characterization of complex balanced states \cite{ge2016ChemicalPhysics,rao2016Phys.Rev.X,yoshimura2021Phys.Rev.Research}, it is still an open problem.
We believe that the Hessian geometric structure can contribute to the resolution of the problem.

\section{Stochastic thermodynamics on Graph}\label{sec:ST}
Finally, we describe how a class of models in stochastic thermodynamics is derived as a special case of our results.

A reversible Markov chain on a graph is given by
\begin{align}
 \frac{\dd p_{i}}{\dd t} = \sum_{j=1}^{N}\left[W_{i\|j}p_{j}- W_{j\|i}p_{i}\right]\label{eq:Markov:chain}
\end{align}
is often employed as a model of stochastic thermodynamics \cite{hill2005,schnakenberg1976Rev.Mod.Phys.,seifert2012Rep.Prog.Phys.,ito2018Phys.Rev.Lett.}. 
Here, $i$ is the index of vertices of the graph, and $N$ is the total number of vertices.
$W_{i\|j}$ is the transition rate from state $j$ to $i$.
Here, the transition from a vertex to itself is usually prohibited, i.e., $W_{i\|i}=0$.
By assuming $W_{i\|j}>0$ for $i\neq j$, the reversible Markov chain can be represented by a complete graph. 
We label the edges by running variable $m$, and determine the orientation of each edge arbitrary.
Let $\IncMatrix$ be the incidence matrix of the oriented graph.
We define functions $h(m)$ and $t(m)$ that return the index of head and tail vertices of the $m$th oriented edge, respectively.
Then, \eqnref{eq:Markov:chain} can be mapped to a chemical rate equation (\eqnref{eq:CRN}) by defining $x_{i}=p_{i}$, $Y=\identityM$, $\stoiMatrix=-Y\IncMatrix$, and 
\begin{align}
    j^{+}_{m}(\Vc{x})&=W_{t(m)\|h(m)}x_{h(m)},\\ j^{-}_{m}(\Vc{x})&=W_{h(m)\|t(m)}x_{t(m)}.
\end{align}
All fluxes are linear in this case.
Because the graph is fully connected, $\IncMatrix$ has the unique left null vector $\Vc{1}$ as $\Vc{1}^{\Transpose}\IncMatrix=\Vc{0}$.
Thus, the system has only one conserved quantity $\Vc{\eta}=\Vc{1}^{\Transpose}\Vc{p}$, which reflects the conservation of total probability, i.e., $\Vc{1}^{\Transpose}\Vc{p}=1$.
Thus $\U=\Vc{1}^{\Transpose}$, and $\Polytope_{\X}(\eta)=\{\Vc{x}|\eta=\U\Vc{x}\}=\{\Vc{x}|\eta=\Vc{1}^{\Transpose}\Vc{x}\}$.
The stoichiometric polytope is nothing but the $N-1$ dimensional simplex of probability distributions if we fix $\eta=1$.
Even if $\IncMatrix$ is not fully connected, we typically assume that the graph is strongly connected to assure that \eqnref{eq:Markov:chain} has a unique and globally stable steady state $\pi$ by the Perron-Frobenius theorem.
In this case, $\IncMatrix$ has the unique left null vector $\Vc{1}$ as well.

The detailed balance condition of \eqnref{eq:Markov:chain} is typically defined as 
\begin{align}
W_{i\|j}\pi_{j}=W_{j\|i}\pi_{i}\quad \mbox{for all $i$ and $j$}.
\end{align}
This is equivalent to the detailed balance condition of the chemical rate equation: $j^{+}_{m}(\Vc{x}_{eq})=j^{-}_{m}(\Vc{x}_{eq})$ for all $m$, i.e., $\Vc{j}(\Vc{x}_{eq})=0$, whereby $\Vc{x}_{eq}=\Vc{\pi}$.
The equilibrium constant characterizing an equilibrium state becomes  $K_{m}=W_{t(m)\|h(m)}/W_{h(m)\|t(m)}$, which also satisfies 
\begin{align}
\ln \Vc{K}=-\IncMatrix^{\Transpose}\ln \Vc{\pi} \label{eq:lnKpi}
\end{align}
as follows from the detailed balance condition.
Thus, the equilibrium \variety can be defined for the Markov chain.
Its explicit parametrization (\eqnref{eq:VarEq}) is given by
\begin{align}
    \Variety_{eq}^{\X}(\Vc{K})&=\left\{\Vc{x}|\Vc{x}=\tilde{\Vc{x}}_{eq}\circ\exp [\Vc{1}\eta^{*}]\right\}.
\end{align}
Here, we used that $\U=\Vc{1}^{\Transpose}$ and that $\tilde{\Vc{x}}_{eq}$ satisfies $\ln \Vc{K}=\stoiMatrix^{\Transpose}\ln\tilde{\Vc{x}}_{eq}=-\IncMatrix^{\Transpose}\ln\tilde{\Vc{x}}_{eq}$.
From \eqnref{eq:lnKpi}, choosing the identification $\tilde{\Vc{x}}_{eq} = \Vc{\pi}$ yields
\begin{align}
    \Variety_{eq}^{\X}(\Vc{K})&=\left\{\Vc{x}|\Vc{x}=\Vc{\pi}\circ\exp [\Vc{1}\eta^{*}]\right\}. 
\end{align}
The \variety $\Variety_{eq}^{\X}(\Vc{K})$ is one-dimensional because the codimension of $\Polytope_{\X}(\eta)$ is one.
The intersection of $\Variety_{eq}^{\X}(\Vc{K})$ and $\Polytope_{\X}(\eta)$ is trivially $\Vc{\pi}$ if $\eta = 1$.
This means that, in the case of stochastic dynamics on a graph, the Hessian structure is not evident because the equilibrium \variety is just one-dimensional and the polytope is always fixed at $\eta=1$, i.e., $\Vc{1}^{\Transpose}\Vc{p}=1$ .
Under this constraint, the potential function is reduced to the conventional Kullback-Leibler divergence, $\varphi(\Vc{p})=\Vc{p}^{\Transpose}\ln (\Vc{p}/\hat{\Vc{p}})$, where we define $\ln \hat{\Vc{p}}\defeq \hat{y}$. 
The Legendre dual of $\Vc{p}$ is $\Vc{y}=\partial \varphi(\Vc{p})=\ln \Vc{p}-\ln \hat{\Vc{p}}$.
Thus, even though $\ln \Vc{p}$ looks as if it is just the logarithm of $\Vc{p}$, the vectors $\Vc{p}$ and $\ln\Vc{p}$ should be discriminated as objects defined on different spaces.
The dual of $\varphi$ is $\varphi^{*}(\Vc{y})=\Vc{1}^{\Transpose}e^{\Vc{y}+\hat{\Vc{y}}}$.
The Bregman divergence is now reduced to the conventional Kullback-Leibler divergence $\KL[\Vc{p}\|\Vc{p}']=\Vc{p}^{\Transpose}\ln(\Vc{p}/\Vc{p}')$.
Even though $\varphi(\Vc{p})$ and $\KL[\Vc{p}\|\Vc{p}']$ are different quantities, both have the same expression as a KL divergence, which can cause confusion.
Finally, the Pythagorean relation does not provide nontrivial information because the polytope $\Polytope_{\X}(\eta)$ is fixed as the probability simplex and the \variety $\Variety_{eq}^{\X}(\Vc{K})$ is one-dimensional.
Thus, the Hessian geometric structure is not fully exhibited under the restricted setting of a Markov chain model.
Nonetheless, the above discussion is beneficial because it allows to discriminate degenerate quantities that have different meanings, such as $\Vc{p}$ and $\ln\Vc{p}$ or $\varphi(\Vc{x})$ and $\KL_{\X}$ and also to point out their locations in dual spaces \cite{ohga2021ArXiv211211008Cond-Mat}.

\section{Discussion}
In this work, we derived the Hessian geometric structure of chemical reaction systems which satisfy the law of mass action. 
When establishing the geometry, a crucial role is played by the fact that the equilibrium and complex balanced states are given by toric varieties. 
A purely thermodynamic argument in \cite{sughiyama2021ArXiv211212403Cond-MatPhysicsphysics} assures, however, that this geometric structure holds for a broader class of chemical reaction systems than those obeying mass action kinetics. 
In the last century, there have been continuous attempts to extend the properties of equilibrium and complex balanced systems beyond the law of mass action in physics and applied mathematics \cite{,avanzini2021J.Chem.Phys.,craciun2019MBE,adamer2020JMathChem}.  
Thus, we expect that our results can provide a new theoretical basis for the development of chemical reaction network theory to such generalized models.

Our current approach relies on the fact that the state space of the system is represented by finite-dimensional vectors: the concentration vector in chemical reaction systems and the probability vector in the Markov chain model. 
However, in reaction systems with a spatially inhomogeneous structure or in general stochastic thermodynamic models, the system is described by a positive or stochastic measure on a continuous space. 
It is an open question whether the structure presented here can be extended to such cases. 
It is naturally expected that information or Hessian geometry on an infinite-dimensional space becomes necessary to address the problem \cite{ay2017,newton2012JournalofFunctionalAnalysis}.

For the theory of information geometry and Hessian geometry \cite{amari2016,shima2007}, chemical reaction systems provide a new and fertile field to apply and develop the theory. 
While major applications of the theory have been restricted to statistics, information science, and other applied mathematical topics, more recently, applications to stochastic thermodynamics have been attempted \cite{crooks2007Phys.Rev.Lett.,ito2018Phys.Rev.Lett.,kolchinsky2021Phys.Rev.X,ohga2021ArXiv211211008Cond-Mat}.
Now, we add chemical reaction systems to the list. 
In addition, from the viewpoint of real algebraic geometry, our results on chemical reaction systems may suggest a way to generalize Birch’s theorem to more general kinetic models \cite{craciun2019MBE}.

The mathematical structure of chemical kinetics and chemical thermodynamics, since their establishment, has been continuously developed within various fields such as physics, applied mathematics, applied chemistry, and systems biology over the past century \cite{beard2008,rao2016Phys.Rev.X,aris1965,feinberg2019,alon2019}. 
However, the developments were mostly separated and shared only within the individual fields. 
In applied mathematics, there is chemical reaction network theory by Feinberg \cite{feinberg2019}, which is based on the work of Aris, Horn, and Jackson \cite{aris1965,horn1972Arch.RationalMech.Anal.a}. 
In real algebraic geometry, chemical reaction systems and toric geometry are becoming important research topics \cite{gopalkrishnan2014SIAMJ.Appl.Dyn.Syst.,craciun2016ArXiv150102860Math,dickenstein2019BullMathBiol,joshi2017SIAMJ.Appl.Dyn.Syst.}. 
In systems biology, a new theory emerged, which connects properties of reaction networks with the network topology \cite{shinar2010Science,okada2016Phys.Rev.Lett.,mochizuki2015JournalofTheoreticalBiology,fiedler2015Math.MethodsAppl.Sci.,hirono2021Phys.Rev.Research}. 
In physics, network thermodynamics by Hill and Schnakenberg \cite{hill2005,schnakenberg1976Rev.Mod.Phys.} and stochastic thermodynamics of chemical reaction theory by Qian and Esposito have been studied \cite{beard2008,rao2016Phys.Rev.X,qian2021}. 
Now, information and Hessian geometry can be added to this variety of applications.
Even though the theories have been developed to explain the same physical object, i.e., a chemical reaction system, the interrelationships between them are not clear yet. 
It will be the next important step to integrate these theories from a unified perspective, which is expected to boost a further development of chemical reaction theory.

We note that a draft on a similar topic appeared on arXiv very recently (27 Dec 2021) \cite{ohga2021ArXiv211213813Cond-MatPhysicsphysicsa}.

\section{Acknowledgement}

This research is supported by JSPS KAKENHI Grant Numbers 19H05799 and 21K21308, and by JST CREST JPMJCR2011 and JPMJCR1927. We thank the members of our lab for the fruitful discussion.


\bibliography{2021_Dimitri.bib,2021_Kobayashi.bib}

\begin{thebibliography}{63}%
\makeatletter
\providecommand \@ifxundefined [1]{%
 \@ifx{#1\undefined}
}%
\providecommand \@ifnum [1]{%
 \ifnum #1\expandafter \@firstoftwo
 \else \expandafter \@secondoftwo
 \fi
}%
\providecommand \@ifx [1]{%
 \ifx #1\expandafter \@firstoftwo
 \else \expandafter \@secondoftwo
 \fi
}%
\providecommand \natexlab [1]{#1}%
\providecommand \enquote  [1]{``#1''}%
\providecommand \bibnamefont  [1]{#1}%
\providecommand \bibfnamefont [1]{#1}%
\providecommand \citenamefont [1]{#1}%
\providecommand \href@noop [0]{\@secondoftwo}%
\providecommand \href [0]{\begingroup \@sanitize@url \@href}%
\providecommand \@href[1]{\@@startlink{#1}\@@href}%
\providecommand \@@href[1]{\endgroup#1\@@endlink}%
\providecommand \@sanitize@url [0]{\catcode `\\12\catcode `\$12\catcode
  `\&12\catcode `\#12\catcode `\^12\catcode `\_12\catcode `\%12\relax}%
\providecommand \@@startlink[1]{}%
\providecommand \@@endlink[0]{}%
\providecommand \url  [0]{\begingroup\@sanitize@url \@url }%
\providecommand \@url [1]{\endgroup\@href {#1}{\urlprefix }}%
\providecommand \urlprefix  [0]{URL }%
\providecommand \Eprint [0]{\href }%
\providecommand \doibase [0]{https://doi.org/}%
\providecommand \selectlanguage [0]{\@gobble}%
\providecommand \bibinfo  [0]{\@secondoftwo}%
\providecommand \bibfield  [0]{\@secondoftwo}%
\providecommand \translation [1]{[#1]}%
\providecommand \BibitemOpen [0]{}%
\providecommand \bibitemStop [0]{}%
\providecommand \bibitemNoStop [0]{.\EOS\space}%
\providecommand \EOS [0]{\spacefactor3000\relax}%
\providecommand \BibitemShut  [1]{\csname bibitem#1\endcsname}%
\let\auto@bib@innerbib\@empty
\bibitem [{\citenamefont {Alon}(2019)}]{alon2019}%
  \BibitemOpen
  \bibfield  {author} {\bibinfo {author} {\bibfnamefont {U.}~\bibnamefont
  {Alon}},\ }\href@noop {} {\emph {\bibinfo {title} {An {{Introduction}} to
  {{Systems Biology}}: {{Design Principles}} of {{Biological Circuits}}}}}\
  (\bibinfo  {publisher} {{CRC Press LLC}},\ \bibinfo {year}
  {2019})\BibitemShut {NoStop}%
\bibitem [{\citenamefont {Mikhailov}\ and\ \citenamefont
  {Ertl}(2017)}]{mikhailov2017}%
  \BibitemOpen
  \bibfield  {author} {\bibinfo {author} {\bibfnamefont {A.~S.}\ \bibnamefont
  {Mikhailov}}\ and\ \bibinfo {author} {\bibfnamefont {G.}~\bibnamefont
  {Ertl}},\ }\href@noop {} {\emph {\bibinfo {title} {Chemical {{Complexity}}:
  {{Self}}-{{Organization Processes}} in {{Molecular Systems}}}}}\ (\bibinfo
  {publisher} {{Springer}},\ \bibinfo {year} {2017})\BibitemShut {NoStop}%
\bibitem [{\citenamefont {Feinberg}(2019)}]{feinberg2019}%
  \BibitemOpen
  \bibfield  {author} {\bibinfo {author} {\bibfnamefont {M.}~\bibnamefont
  {Feinberg}},\ }\href@noop {} {\emph {\bibinfo {title} {Foundations of
  {{Chemical Reaction Network Theory}}}}}\ (\bibinfo  {publisher}
  {{Springer}},\ \bibinfo {year} {2019})\BibitemShut {NoStop}%
\bibitem [{\citenamefont {Murray}(2011)}]{murray2011}%
  \BibitemOpen
  \bibfield  {author} {\bibinfo {author} {\bibfnamefont {J.~D.}\ \bibnamefont
  {Murray}},\ }\href@noop {} {\emph {\bibinfo {title} {Mathematical
  {{Biology}}: {{I}}. {{An Introduction}}}}}\ (\bibinfo  {publisher} {{Springer
  Science \& Business Media}},\ \bibinfo {year} {2011})\BibitemShut {NoStop}%
\bibitem [{\citenamefont {Epstein}\ and\ \citenamefont
  {Pojman}(1998)}]{epstein1998}%
  \BibitemOpen
  \bibfield  {author} {\bibinfo {author} {\bibfnamefont {I.~R.}\ \bibnamefont
  {Epstein}}\ and\ \bibinfo {author} {\bibfnamefont {J.~A.}\ \bibnamefont
  {Pojman}},\ }\href {https://doi.org/10.1093/oso/9780195096705.001.0001}
  {\emph {\bibinfo {title} {An Introduction to Nonlinear Chemical
  Dynamics: Oscillations, Waves, Patterns, and Chaos}}},\
  Topics in {{Physical Chemistry}}\ (\bibinfo  {publisher} {{Oxford University
  Press}},\ \bibinfo {address} {{New York}},\ \bibinfo {year}
  {1998})\BibitemShut {NoStop}%
\bibitem [{\citenamefont {Beard}\ and\ \citenamefont {Qian}(2008)}]{beard2008}%
  \BibitemOpen
  \bibfield  {author} {\bibinfo {author} {\bibfnamefont {D.~A.}\ \bibnamefont
  {Beard}}\ and\ \bibinfo {author} {\bibfnamefont {H.}~\bibnamefont {Qian}},\
  }\href {https://doi.org/10.1017/CBO9780511803345} {\emph {\bibinfo {title}
  {Chemical {{Biophysics}}: {{Quantitative Analysis}} of {{Cellular
  Systems}}}}},\ Cambridge {{Texts}} in {{Biomedical Engineering}}\ (\bibinfo
  {publisher} {{Cambridge University Press}},\ \bibinfo {address}
  {{Cambridge}},\ \bibinfo {year} {2008})\BibitemShut {NoStop}%
\bibitem [{\citenamefont
  {Wegscheider}(1902)}]{wegscheider1902Z.FuerPhys.Chem.}%
  \BibitemOpen
  \bibfield  {author} {\bibinfo {author} {\bibfnamefont {R.}~\bibnamefont
  {Wegscheider}},\ }\bibfield  {title} {\bibinfo {title} {{\"Uber simultane
  Gleichgewichte und die Beziehungen zwischen Thermodynamik und
  Reaktionskinetik homogener Systeme}},\ }\href
  {https://doi.org/10.1515/zpch-1902-3919} {\bibfield  {journal} {\bibinfo
  {journal} {Zeitschrift f\"ur Physikalische Chemie}\ }\textbf {\bibinfo
  {volume} {39U}},\ \bibinfo {pages} {257} (\bibinfo {year}
  {1902})}\BibitemShut {NoStop}%
\bibitem [{\citenamefont {Hill}(2005)}]{hill2005}%
  \BibitemOpen
  \bibfield  {author} {\bibinfo {author} {\bibfnamefont {T.~L.}\ \bibnamefont
  {Hill}},\ }\href@noop {} {\emph {\bibinfo {title} {Free {{Energy
  Transduction}} and {{Biochemical Cycle Kinetics}}}}}\ (\bibinfo  {publisher}
  {{Courier Corporation}},\ \bibinfo {year} {2005})\BibitemShut {NoStop}%
\bibitem [{\citenamefont {Hill}(1966)}]{hill1966JournalofTheoreticalBiology}%
  \BibitemOpen
  \bibfield  {author} {\bibinfo {author} {\bibfnamefont {T.~L.}\ \bibnamefont
  {Hill}},\ }\bibfield  {title} {\bibinfo {title} {Studies in irreversible
  thermodynamics {{IV}}. diagrammatic representation of steady state fluxes for
  unimolecular systems},\ }\href {https://doi.org/10.1016/0022-5193(66)90137-8}
  {\bibfield  {journal} {\bibinfo  {journal} {Journal of Theoretical Biology}\
  }\textbf {\bibinfo {volume} {10}},\ \bibinfo {pages} {442} (\bibinfo {year}
  {1966})}\BibitemShut {NoStop}%
\bibitem [{\citenamefont {Schnakenberg}(1976)}]{schnakenberg1976Rev.Mod.Phys.}%
  \BibitemOpen
  \bibfield  {author} {\bibinfo {author} {\bibfnamefont {J.}~\bibnamefont
  {Schnakenberg}},\ }\bibfield  {title} {\bibinfo {title} {Network theory of
  microscopic and macroscopic behavior of master equation systems},\ }\href
  {https://doi.org/10.1103/RevModPhys.48.571} {\bibfield  {journal} {\bibinfo
  {journal} {Reviews of Modern Physics}\ }\textbf {\bibinfo {volume} {48}},\
  \bibinfo {pages} {571} (\bibinfo {year} {1976})}\BibitemShut {NoStop}%
\bibitem [{\citenamefont {Polettini}\ and\ \citenamefont
  {Esposito}(2014)}]{polettini2014J.Chem.Phys.}%
  \BibitemOpen
  \bibfield  {author} {\bibinfo {author} {\bibfnamefont {M.}~\bibnamefont
  {Polettini}}\ and\ \bibinfo {author} {\bibfnamefont {M.}~\bibnamefont
  {Esposito}},\ }\bibfield  {title} {\bibinfo {title} {Irreversible
  thermodynamics of open chemical networks. {{I}}. {{Emergent}} cycles and
  broken conservation laws},\ }\href {https://doi.org/10.1063/1.4886396}
  {\bibfield  {journal} {\bibinfo  {journal} {The Journal of Chemical Physics}\
  }\textbf {\bibinfo {volume} {141}},\ \bibinfo {pages} {024117} (\bibinfo
  {year} {2014})}\BibitemShut {NoStop}%
\bibitem [{\citenamefont {Rao}\ and\ \citenamefont
  {Esposito}(2016)}]{rao2016Phys.Rev.X}%
  \BibitemOpen
  \bibfield  {author} {\bibinfo {author} {\bibfnamefont {R.}~\bibnamefont
  {Rao}}\ and\ \bibinfo {author} {\bibfnamefont {M.}~\bibnamefont {Esposito}},\
  }\bibfield  {title} {\bibinfo {title} {Nonequilibrium {{Thermodynamics}} of
  {{Chemical Reaction Networks}}: {{Wisdom}} from {{Stochastic
  Thermodynamics}}},\ }\href {https://doi.org/10.1103/PhysRevX.6.041064}
  {\bibfield  {journal} {\bibinfo  {journal} {Physical Review X}\ }\textbf
  {\bibinfo {volume} {6}},\ \bibinfo {pages} {041064} (\bibinfo {year}
  {2016})}\BibitemShut {NoStop}%
\bibitem [{\citenamefont {Rao}\ and\ \citenamefont
  {Esposito}(2018{\natexlab{a}})}]{rao2018J.Chem.Phys.}%
  \BibitemOpen
  \bibfield  {author} {\bibinfo {author} {\bibfnamefont {R.}~\bibnamefont
  {Rao}}\ and\ \bibinfo {author} {\bibfnamefont {M.}~\bibnamefont {Esposito}},\
  }\bibfield  {title} {\bibinfo {title} {Conservation laws and work fluctuation
  relations in chemical reaction networks},\ }\href
  {https://doi.org/10.1063/1.5042253} {\bibfield  {journal} {\bibinfo
  {journal} {The Journal of Chemical Physics}\ }\textbf {\bibinfo {volume}
  {149}},\ \bibinfo {pages} {245101} (\bibinfo {year}
  {2018}{\natexlab{a}})}\BibitemShut {NoStop}%
\bibitem [{\citenamefont {Rao}\ and\ \citenamefont
  {Esposito}(2018{\natexlab{b}})}]{rao2018NewJ.Phys.}%
  \BibitemOpen
  \bibfield  {author} {\bibinfo {author} {\bibfnamefont {R.}~\bibnamefont
  {Rao}}\ and\ \bibinfo {author} {\bibfnamefont {M.}~\bibnamefont {Esposito}},\
  }\bibfield  {title} {\bibinfo {title} {Conservation laws shape dissipation},\
  }\href {https://doi.org/10.1088/1367-2630/aaa15f} {\bibfield  {journal}
  {\bibinfo  {journal} {New Journal of Physics}\ }\textbf {\bibinfo {volume}
  {20}},\ \bibinfo {pages} {023007} (\bibinfo {year}
  {2018}{\natexlab{b}})}\BibitemShut {NoStop}%
\bibitem [{\citenamefont {Avanzini}\ \emph {et~al.}(2021)\citenamefont
  {Avanzini}, \citenamefont {Penocchio}, \citenamefont {Falasco},\ and\
  \citenamefont {Esposito}}]{avanzini2021J.Chem.Phys.}%
  \BibitemOpen
  \bibfield  {author} {\bibinfo {author} {\bibfnamefont {F.}~\bibnamefont
  {Avanzini}}, \bibinfo {author} {\bibfnamefont {E.}~\bibnamefont {Penocchio}},
  \bibinfo {author} {\bibfnamefont {G.}~\bibnamefont {Falasco}},\ and\ \bibinfo
  {author} {\bibfnamefont {M.}~\bibnamefont {Esposito}},\ }\bibfield  {title}
  {\bibinfo {title} {Nonequilibrium thermodynamics of non-ideal chemical
  reaction networks},\ }\href {https://doi.org/10.1063/5.0041225} {\bibfield
  {journal} {\bibinfo  {journal} {The Journal of Chemical Physics}\ }\textbf
  {\bibinfo {volume} {154}},\ \bibinfo {pages} {094114} (\bibinfo {year}
  {2021})}\BibitemShut {NoStop}%
\bibitem [{\citenamefont {Ge}\ and\ \citenamefont
  {Qian}(2016)}]{ge2016ChemicalPhysics}%
  \BibitemOpen
  \bibfield  {author} {\bibinfo {author} {\bibfnamefont {H.}~\bibnamefont
  {Ge}}\ and\ \bibinfo {author} {\bibfnamefont {H.}~\bibnamefont {Qian}},\
  }\bibfield  {title} {\bibinfo {title} {Nonequilibrium thermodynamic formalism
  of nonlinear chemical reaction systems with
  {{Waage}}\textendash{{Guldberg}}'s law of mass action},\ }\href
  {https://doi.org/10.1016/j.chemphys.2016.03.026} {\bibfield  {journal}
  {\bibinfo  {journal} {Chemical Physics}\ }\textbf {\bibinfo {volume} {472}},\
  \bibinfo {pages} {241} (\bibinfo {year} {2016})}\BibitemShut {NoStop}%
\bibitem [{\citenamefont {Schmiedl}\ and\ \citenamefont
  {Seifert}(2007)}]{schmiedl2007J.Chem.Phys.}%
  \BibitemOpen
  \bibfield  {author} {\bibinfo {author} {\bibfnamefont {T.}~\bibnamefont
  {Schmiedl}}\ and\ \bibinfo {author} {\bibfnamefont {U.}~\bibnamefont
  {Seifert}},\ }\bibfield  {title} {\bibinfo {title} {Stochastic thermodynamics
  of chemical reaction networks},\ }\href {https://doi.org/10.1063/1.2428297}
  {\bibfield  {journal} {\bibinfo  {journal} {The Journal of Chemical Physics}\
  }\textbf {\bibinfo {volume} {126}},\ \bibinfo {pages} {044101} (\bibinfo
  {year} {2007})}\BibitemShut {NoStop}%
\bibitem [{\citenamefont {Horn}\ and\ \citenamefont
  {Jackson}(1972)}]{horn1972Arch.RationalMech.Anal.a}%
  \BibitemOpen
  \bibfield  {author} {\bibinfo {author} {\bibfnamefont {F.}~\bibnamefont
  {Horn}}\ and\ \bibinfo {author} {\bibfnamefont {R.}~\bibnamefont {Jackson}},\
  }\bibfield  {title} {\bibinfo {title} {General mass action kinetics},\ }\href
  {https://doi.org/10.1007/BF00251225} {\bibfield  {journal} {\bibinfo
  {journal} {Archive for Rational Mechanics and Analysis}\ }\textbf {\bibinfo
  {volume} {47}},\ \bibinfo {pages} {81} (\bibinfo {year} {1972})}\BibitemShut
  {NoStop}%
\bibitem [{\citenamefont {Craciun}\ \emph {et~al.}(2009)\citenamefont
  {Craciun}, \citenamefont {Dickenstein}, \citenamefont {Shiu},\ and\
  \citenamefont {Sturmfels}}]{craciun2009JournalofSymbolicComputation}%
  \BibitemOpen
  \bibfield  {author} {\bibinfo {author} {\bibfnamefont {G.}~\bibnamefont
  {Craciun}}, \bibinfo {author} {\bibfnamefont {A.}~\bibnamefont
  {Dickenstein}}, \bibinfo {author} {\bibfnamefont {A.}~\bibnamefont {Shiu}},\
  and\ \bibinfo {author} {\bibfnamefont {B.}~\bibnamefont {Sturmfels}},\
  }\bibfield  {title} {\bibinfo {title} {Toric dynamical systems},\ }\href
  {https://doi.org/10.1016/j.jsc.2008.08.006} {\bibfield  {journal} {\bibinfo
  {journal} {Journal of Symbolic Computation}\ }\bibinfo {series} {In
  Memoriam Karin Gatermann},\ \textbf {\bibinfo volume 44},\ \bibinfo
  {pages} {1551} (\bibinfo {year} {2009})}\BibitemShut {NoStop}%
\bibitem [{\citenamefont {Craciun}\ \emph {et~al.}(2019)\citenamefont
  {Craciun}, \citenamefont {Muller}, \citenamefont {Pantea}, \citenamefont
  {Yu}, \citenamefont {Craciun}, \citenamefont {Muller}, \citenamefont
  {Pantea},\ and\ \citenamefont {Yu}}]{craciun2019MBE}%
  \BibitemOpen
  \bibfield  {author} {\bibinfo {author} {\bibfnamefont {G.}~\bibnamefont
  {Craciun}}, \bibinfo {author} {\bibfnamefont {S.}~\bibnamefont {Muller}},
  \bibinfo {author} {\bibfnamefont {C.}~\bibnamefont {Pantea}}, \bibinfo
  {author} {\bibfnamefont {P.~Y.}\ \bibnamefont {Yu}}, \bibinfo {author}
  {\bibfnamefont {G.}~\bibnamefont {Craciun}}, \bibinfo {author} {\bibfnamefont
  {S.}~\bibnamefont {Muller}}, \bibinfo {author} {\bibfnamefont
  {C.}~\bibnamefont {Pantea}},\ and\ \bibinfo {author} {\bibfnamefont {P.~Y.}\
  \bibnamefont {Yu}},\ }\bibfield  {title} {\bibinfo {title} {A generalization
  of {{Birchs}} theorem and vertex-balanced steady states for generalized
  mass-action systems},\ }\href {https://doi.org/10.3934/mbe.2019417}
  {\bibfield  {journal} {\bibinfo  {journal} {Mathematical Biosciences and
  Engineering}\ }\textbf {\bibinfo {volume} {16}},\ \bibinfo {pages} {8243}
  (\bibinfo {year} {2019})}\BibitemShut {NoStop}%
\bibitem [{\citenamefont {Craciun}\ and\ \citenamefont
  {Sorea}(2020)}]{craciun2020ArXiv200811468Math}%
  \BibitemOpen
  \bibfield  {author} {\bibinfo {author} {\bibfnamefont {G.}~\bibnamefont
  {Craciun}}\ and\ \bibinfo {author} {\bibfnamefont {M.-S.}\ \bibnamefont
  {Sorea}},\ }\bibfield  {title} {\bibinfo {title} {The structure of the moduli
  spaces of toric dynamical systems},\ }\href@noop {} {\bibfield  {journal}
  {\bibinfo  {journal} {arXiv:2008.11468 [math]}\ } (\bibinfo {year} {2020})},\
  \Eprint {https://arxiv.org/abs/2008.11468} {arXiv:2008.11468 [math]}
  \BibitemShut {NoStop}%
\bibitem [{\citenamefont {{van der Schaft}}\ \emph {et~al.}(2013)\citenamefont
  {{van der Schaft}}, \citenamefont {Rao},\ and\ \citenamefont
  {Jayawardhana}}]{vanderschaft2013SIAMJ.Appl.Math.}%
  \BibitemOpen
  \bibfield  {author} {\bibinfo {author} {\bibfnamefont {A.}~\bibnamefont {{van
  der Schaft}}}, \bibinfo {author} {\bibfnamefont {S.}~\bibnamefont {Rao}},\
  and\ \bibinfo {author} {\bibfnamefont {B.}~\bibnamefont {Jayawardhana}},\
  }\bibfield  {title} {\bibinfo {title} {On the {{Mathematical Structure}} of
  {{Balanced Chemical Reaction Networks Governed}} by {{Mass Action
  Kinetics}}},\ }\href {https://doi.org/10.1137/11085431X} {\bibfield
  {journal} {\bibinfo  {journal} {SIAM Journal on Applied Mathematics}\
  }\textbf {\bibinfo {volume} {73}},\ \bibinfo {pages} {953} (\bibinfo {year}
  {2013})}\BibitemShut {NoStop}%
\bibitem [{\citenamefont {Okada}\ and\ \citenamefont
  {Mochizuki}(2016)}]{okada2016Phys.Rev.Lett.}%
  \BibitemOpen
  \bibfield  {author} {\bibinfo {author} {\bibfnamefont {T.}~\bibnamefont
  {Okada}}\ and\ \bibinfo {author} {\bibfnamefont {A.}~\bibnamefont
  {Mochizuki}},\ }\bibfield  {title} {\bibinfo {title} {Law of {{Localization}}
  in {{Chemical Reaction Networks}}},\ }\href
  {https://doi.org/10.1103/PhysRevLett.117.048101} {\bibfield  {journal}
  {\bibinfo  {journal} {Physical Review Letters}\ }\textbf {\bibinfo {volume}
  {117}},\ \bibinfo {pages} {048101} (\bibinfo {year} {2016})}\BibitemShut
  {NoStop}%
\bibitem [{\citenamefont {Mochizuki}\ and\ \citenamefont
  {Fiedler}(2015)}]{mochizuki2015JournalofTheoreticalBiology}%
  \BibitemOpen
  \bibfield  {author} {\bibinfo {author} {\bibfnamefont {A.}~\bibnamefont
  {Mochizuki}}\ and\ \bibinfo {author} {\bibfnamefont {B.}~\bibnamefont
  {Fiedler}},\ }\bibfield  {title} {\bibinfo {title} {Sensitivity of chemical
  reaction networks: {{A}} structural approach. 1. {{Examples}} and the carbon
  metabolic network},\ }\href {https://doi.org/10.1016/j.jtbi.2014.10.025}
  {\bibfield  {journal} {\bibinfo  {journal} {Journal of Theoretical Biology}\
  }\textbf {\bibinfo {volume} {367}},\ \bibinfo {pages} {189} (\bibinfo {year}
  {2015})}\BibitemShut {NoStop}%
\bibitem [{\citenamefont {Fiedler}\ and\ \citenamefont
  {Mochizuki}(2015)}]{fiedler2015Math.MethodsAppl.Sci.}%
  \BibitemOpen
  \bibfield  {author} {\bibinfo {author} {\bibfnamefont {B.}~\bibnamefont
  {Fiedler}}\ and\ \bibinfo {author} {\bibfnamefont {A.}~\bibnamefont
  {Mochizuki}},\ }\bibfield  {title} {\bibinfo {title} {Sensitivity of chemical
  reaction networks: A structural approach. 2. {{Regular}} monomolecular
  systems},\ }\href {https://doi.org/10.1002/mma.3436} {\bibfield  {journal}
  {\bibinfo  {journal} Mathematical Methods in the Applied Sciences\ }\textbf
  {\bibinfo {volume} {38}},\ \bibinfo {pages} {3519} (\bibinfo {year}
  {2015})}\BibitemShut {NoStop}%
\bibitem [{\citenamefont {Shinar}\ and\ \citenamefont
  {Feinberg}(2010)}]{shinar2010Science}%
  \BibitemOpen
  \bibfield  {author} {\bibinfo {author} {\bibfnamefont {G.}~\bibnamefont
  {Shinar}}\ and\ \bibinfo {author} {\bibfnamefont {M.}~\bibnamefont
  {Feinberg}},\ }\bibfield  {title} {\bibinfo {title} {Structural {{Sources}}
  of {{Robustness}} in {{Biochemical Reaction Networks}}},\ }\bibfield
  {journal} {\bibinfo  {journal} {Science}\ }\href
  {https://doi.org/10.1126/science.1183372} {10.1126/science.1183372} (\bibinfo
  {year} {2010})\BibitemShut {NoStop}%
\bibitem [{\citenamefont {Araujo}\ and\ \citenamefont
  {Liotta}(2018)}]{araujo2018NatCommun}%
  \BibitemOpen
  \bibfield  {author} {\bibinfo {author} {\bibfnamefont {R.~P.}\ \bibnamefont
  {Araujo}}\ and\ \bibinfo {author} {\bibfnamefont {L.~A.}\ \bibnamefont
  {Liotta}},\ }\bibfield  {title} {\bibinfo {title} {The topological
  requirements for robust perfect adaptation in networks of any size},\ }\href
  {https://doi.org/10.1038/s41467-018-04151-6} {\bibfield  {journal} {\bibinfo
  {journal} {Nature Communications}\ }\textbf {\bibinfo {volume} {9}},\
  \bibinfo {pages} {1757} (\bibinfo {year} {2018})}\BibitemShut {NoStop}%
\bibitem [{\citenamefont {Hirono}\ \emph {et~al.}(2021)\citenamefont {Hirono},
  \citenamefont {Okada}, \citenamefont {Miyazaki},\ and\ \citenamefont
  {Hidaka}}]{hirono2021Phys.Rev.Research}%
  \BibitemOpen
  \bibfield  {author} {\bibinfo {author} {\bibfnamefont {Y.}~\bibnamefont
  {Hirono}}, \bibinfo {author} {\bibfnamefont {T.}~\bibnamefont {Okada}},
  \bibinfo {author} {\bibfnamefont {H.}~\bibnamefont {Miyazaki}},\ and\
  \bibinfo {author} {\bibfnamefont {Y.}~\bibnamefont {Hidaka}},\ }\bibfield
  {title} {\bibinfo {title} {Structural reduction of chemical reaction networks
  based on topology},\ }\href
  {https://doi.org/10.1103/PhysRevResearch.3.043123} {\bibfield  {journal}
  {\bibinfo  {journal} {Physical Review Research}\ }\textbf {\bibinfo {volume}
  {3}},\ \bibinfo {pages} {043123} (\bibinfo {year} {2021})},\ \Eprint
  {https://arxiv.org/abs/2102.07687} {arXiv:2102.07687} \BibitemShut {NoStop}%
\bibitem [{\citenamefont {Gopalkrishnan}\ \emph {et~al.}(2014)\citenamefont
  {Gopalkrishnan}, \citenamefont {Miller},\ and\ \citenamefont
  {Shiu}}]{gopalkrishnan2014SIAMJ.Appl.Dyn.Syst.}%
  \BibitemOpen
  \bibfield  {author} {\bibinfo {author} {\bibfnamefont {M.}~\bibnamefont
  {Gopalkrishnan}}, \bibinfo {author} {\bibfnamefont {E.}~\bibnamefont
  {Miller}},\ and\ \bibinfo {author} {\bibfnamefont {A.}~\bibnamefont {Shiu}},\
  }\bibfield  {title} {\bibinfo {title} {A {{Geometric Approach}} to the
  {{Global Attractor Conjecture}}},\ }\href {https://doi.org/10.1137/130928170}
  {\bibfield  {journal} {\bibinfo  {journal} {SIAM Journal on Applied Dynamical
  Systems}\ }\textbf {\bibinfo {volume} {13}},\ \bibinfo {pages} {758}
  (\bibinfo {year} {2014})}\BibitemShut {NoStop}%
\bibitem [{\citenamefont {Craciun}(2016)}]{craciun2016ArXiv150102860Math}%
  \BibitemOpen
  \bibfield  {author} {\bibinfo {author} {\bibfnamefont {G.}~\bibnamefont
  {Craciun}},\ }\bibfield  {title} {\bibinfo {title} {Toric {{Differential
  Inclusions}} and a {{Proof}} of the {{Global Attractor Conjecture}}},\
  }\href@noop {} {\bibfield  {journal} {\bibinfo  {journal} {arXiv:1501.02860
  [math]}\ } (\bibinfo {year} {2016})},\ \Eprint
  {https://arxiv.org/abs/1501.02860} {arXiv:1501.02860 [math]} \BibitemShut
  {NoStop}%
\bibitem [{\citenamefont {Dickenstein}\ \emph {et~al.}(2019)\citenamefont
  {Dickenstein}, \citenamefont {Mill{\'a}n}, \citenamefont {Shiu},\ and\
  \citenamefont {Tang}}]{dickenstein2019BullMathBiol}%
  \BibitemOpen
  \bibfield  {author} {\bibinfo {author} {\bibfnamefont {A.}~\bibnamefont
  {Dickenstein}}, \bibinfo {author} {\bibfnamefont {M.~P.}\ \bibnamefont
  {Mill{\'a}n}}, \bibinfo {author} {\bibfnamefont {A.}~\bibnamefont {Shiu}},\
  and\ \bibinfo {author} {\bibfnamefont {X.}~\bibnamefont {Tang}},\ }\bibfield
  {title} {\bibinfo {title} {Multistationarity in {{Structured Reaction
  Networks}}},\ }\href {https://doi.org/10.1007/s11538-019-00572-6} {\bibfield
  {journal} {\bibinfo  {journal} {Bulletin of Mathematical Biology}\ }\textbf
  {\bibinfo {volume} {81}},\ \bibinfo {pages} {1527} (\bibinfo {year}
  {2019})}\BibitemShut {NoStop}%
\bibitem [{\citenamefont {Joshi}\ and\ \citenamefont
  {Shiu}(2017)}]{joshi2017SIAMJ.Appl.Dyn.Syst.}%
  \BibitemOpen
  \bibfield  {author} {\bibinfo {author} {\bibfnamefont {B.}~\bibnamefont
  {Joshi}}\ and\ \bibinfo {author} {\bibfnamefont {A.}~\bibnamefont {Shiu}},\
  }\bibfield  {title} {\bibinfo {title} {Which {{Small Reaction Networks Are
  Multistationary}}?},\ }\href {https://doi.org/10.1137/16M1069705} {\bibfield
  {journal} {\bibinfo  {journal} {SIAM Journal on Applied Dynamical Systems}\
  }\textbf {\bibinfo {volume} {16}},\ \bibinfo {pages} {802} (\bibinfo {year}
  {2017})}\BibitemShut {NoStop}%
\bibitem [{\citenamefont {{Otero-Muras}}\ \emph {et~al.}(2017)\citenamefont
  {{Otero-Muras}}, \citenamefont {Yordanov},\ and\ \citenamefont
  {Stelling}}]{otero-muras2017PLOSComputationalBiology}%
  \BibitemOpen
  \bibfield  {author} {\bibinfo {author} {\bibfnamefont {I.}~\bibnamefont
  {{Otero-Muras}}}, \bibinfo {author} {\bibfnamefont {P.}~\bibnamefont
  {Yordanov}},\ and\ \bibinfo {author} {\bibfnamefont {J.}~\bibnamefont
  {Stelling}},\ }\bibfield  {title} {\bibinfo {title} {Chemical {{Reaction
  Network Theory}} elucidates sources of multistability in interferon
  signaling},\ }\href {https://doi.org/10.1371/journal.pcbi.1005454} {\bibfield
   {journal} {\bibinfo  {journal} {PLOS Computational Biology}\ }\textbf
  {\bibinfo {volume} {13}},\ \bibinfo {pages} {e1005454} (\bibinfo {year}
  {2017})}\BibitemShut {NoStop}%
\bibitem [{\citenamefont {Sughiyama}\ \emph {et~al.}(2021)\citenamefont
  {Sughiyama}, \citenamefont {Loutchko}, \citenamefont {Kamimura},\ and\
  \citenamefont
  {Kobayashi}}]{sughiyama2021ArXiv211212403Cond-MatPhysicsphysics}%
  \BibitemOpen
  \bibfield  {author} {\bibinfo {author} {\bibfnamefont {Y.}~\bibnamefont
  {Sughiyama}}, \bibinfo {author} {\bibfnamefont {D.}~\bibnamefont {Loutchko}},
  \bibinfo {author} {\bibfnamefont {A.}~\bibnamefont {Kamimura}},\ and\
  \bibinfo {author} {\bibfnamefont {T.~J.}\ \bibnamefont {Kobayashi}},\
  }\bibfield  {title} {\bibinfo {title} {A {{Hessian Geometric Structure}} of
  {{Chemical Thermodynamic Systems}} with {{Stoichiometric Constraints}}},\
  }\href@noop {} {\bibfield  {journal} {\bibinfo  {journal} {arXiv:2112.12403
  [cond-mat, physics:physics]}\ } (\bibinfo {year} {2021})},\ \Eprint
  {https://arxiv.org/abs/2112.12403} {arXiv:2112.12403 [cond-mat,
  physics:physics]} \BibitemShut {NoStop}%
\bibitem [{\citenamefont {Yoshimura}\ and\ \citenamefont
  {Ito}(2021)}]{yoshimura2021Phys.Rev.Research}%
  \BibitemOpen
  \bibfield  {author} {\bibinfo {author} {\bibfnamefont {K.}~\bibnamefont
  {Yoshimura}}\ and\ \bibinfo {author} {\bibfnamefont {S.}~\bibnamefont
  {Ito}},\ }\bibfield  {title} {\bibinfo {title} {Information geometric
  inequalities of chemical thermodynamics},\ }\href
  {https://doi.org/10.1103/PhysRevResearch.3.013175} {\bibfield  {journal}
  {\bibinfo  {journal} {Physical Review Research}\ }\textbf {\bibinfo {volume}
  {3}},\ \bibinfo {pages} {013175} (\bibinfo {year} {2021})}\BibitemShut
  {NoStop}%
\bibitem [{\citenamefont {Shima}(2007)}]{shima2007}%
  \BibitemOpen
  \bibfield  {author} {\bibinfo {author} {\bibfnamefont {H.}~\bibnamefont
  {Shima}},\ }\href@noop {} {\emph {\bibinfo {title} {The {{Geometry}} of
  {{Hessian Structures}}}}}\ (\bibinfo  {publisher} {{World Scientific}},\
  \bibinfo {year} {2007})\BibitemShut {NoStop}%
\bibitem [{\citenamefont {Amari}(2016)}]{amari2016}%
  \BibitemOpen
  \bibfield  {author} {\bibinfo {author} {\bibfnamefont {S.-i.}\ \bibnamefont
  {Amari}},\ }\href@noop {} {\emph {\bibinfo {title} {Information {{Geometry}}
  and {{Its Applications}}}}}\ (\bibinfo  {publisher} {{Springer}},\ \bibinfo
  {year} {2016})\BibitemShut {NoStop}%
\bibitem [{\citenamefont {Amari}(1982)}]{amari1982Ann.Stat.}%
  \BibitemOpen
  \bibfield  {author} {\bibinfo {author} {\bibfnamefont {S.-I.}\ \bibnamefont
  {Amari}},\ }\bibfield  {title} {\bibinfo {title} {Differential {{Geometry}}
  of {{Curved Exponential Families}}-{{Curvatures}} and {{Information Loss}}},\
  }\href@noop {} {\bibfield  {journal} {\bibinfo  {journal} {The Annals of
  Statistics}\ }\textbf {\bibinfo {volume} {10}},\ \bibinfo {pages} {357}
  (\bibinfo {year} {1982})}\BibitemShut {NoStop}%
\bibitem [{\citenamefont
  {Amari}(1985)}]{amari1985Differential-GeometricalMethodsinStatistics}%
  \BibitemOpen
  \bibfield  {author} {\bibinfo {author} {\bibfnamefont {S.-i.}\ \bibnamefont
  {Amari}},\ }\bibfield  {title} {\bibinfo {title} {Differential {{Geometry}}
  of {{Statistical Models}}},\ }in\ \href
  {https://doi.org/10.1007/978-1-4612-5056-2_2} {\emph {\bibinfo {booktitle}
  {Differential-{{Geometrical Methods}} in {{Statistics}}}}},\ \bibinfo {series
  and number} {Lecture {{Notes}} in {{Statistics}}},\ \bibinfo {editor} {edited
  by\ \bibinfo {editor} {\bibfnamefont {S.-i.}\ \bibnamefont {Amari}}}\
  (\bibinfo  {publisher} {{Springer}},\ \bibinfo {address} {{New York, NY}},\
  \bibinfo {year} {1985})\ pp.\ \bibinfo {pages} {11--65}\BibitemShut {NoStop}%
\bibitem [{\citenamefont {{Barndorff-Nielsen}}\ \emph
  {et~al.}(1986)\citenamefont {{Barndorff-Nielsen}}, \citenamefont {Cox},\ and\
  \citenamefont {Reid}}]{barndorff-nielsen1986Int.Stat.Rev.Rev.Int.Stat.}%
  \BibitemOpen
  \bibfield  {author} {\bibinfo {author} {\bibfnamefont {O.~E.}\ \bibnamefont
  {{Barndorff-Nielsen}}}, \bibinfo {author} {\bibfnamefont {D.~R.}\
  \bibnamefont {Cox}},\ and\ \bibinfo {author} {\bibfnamefont {N.}~\bibnamefont
  {Reid}},\ }\bibfield  {title} {\bibinfo {title} {The {{Role}} of
  {{Differential Geometry}} in {{Statistical Theory}}},\ }\href
  {https://doi.org/10.2307/1403260} {\bibfield  {journal} {\bibinfo  {journal}
  International Statistical Review / Revue Internationale de Statistique\
  }\textbf {\bibinfo {volume} {54}},\ \bibinfo {pages} {83} (\bibinfo {year}
  {1986})}\BibitemShut {NoStop}%
\bibitem [{\citenamefont {Amari}\ and\ \citenamefont
  {Nagaoka}(2000)}]{amari2000}%
  \BibitemOpen
  \bibfield  {author} {\bibinfo {author} {\bibfnamefont {S.-i.}\ \bibnamefont
  {Amari}}\ and\ \bibinfo {author} {\bibfnamefont {H.}~\bibnamefont
  {Nagaoka}},\ }\href@noop {} {\emph {\bibinfo {title} {Methods of
  {{Information Geometry}}}}}\ (\bibinfo  {publisher} {{American Mathematical
  Soc.}},\ \bibinfo {year} {2000})\BibitemShut {NoStop}%
\bibitem [{\citenamefont {Amari}\ and\ \citenamefont
  {Han}(1989)}]{amari1989IEEETrans.Inf.Theory}%
  \BibitemOpen
  \bibfield  {author} {\bibinfo {author} {\bibfnamefont {S.-I.}\ \bibnamefont
  {Amari}}\ and\ \bibinfo {author} {\bibfnamefont {T.}~\bibnamefont {Han}},\
  }\bibfield  {title} {\bibinfo {title} {Statistical inference under
  multiterminal rate restrictions: A differential geometric approach},\ }\href
  {https://doi.org/10.1109/18.32118} {\bibfield  {journal} {\bibinfo  {journal}
  {IEEE Transactions on Information Theory}\ }\textbf {\bibinfo {volume}
  {35}},\ \bibinfo {pages} {217} (\bibinfo {year} {1989})}\BibitemShut
  {NoStop}%
\bibitem [{\citenamefont {Okamoto}\ \emph {et~al.}(1991)\citenamefont
  {Okamoto}, \citenamefont {Amari},\ and\ \citenamefont
  {Takeuchi}}]{okamoto1991Ann.Stat.}%
  \BibitemOpen
  \bibfield  {author} {\bibinfo {author} {\bibfnamefont {I.}~\bibnamefont
  {Okamoto}}, \bibinfo {author} {\bibfnamefont {S.-I.}\ \bibnamefont {Amari}},\
  and\ \bibinfo {author} {\bibfnamefont {K.}~\bibnamefont {Takeuchi}},\
  }\bibfield  {title} {\bibinfo {title} {Asymptotic {{Theory}} of {{Sequential
  Estimation}}: {{Differential Geometrical Approach}}},\ }\href
  {https://doi.org/10.1214/aos/1176348131} {\bibfield  {journal} {\bibinfo
  {journal} {The Annals of Statistics}\ }\textbf {\bibinfo {volume} {19}},\
  \bibinfo {pages} {961} (\bibinfo {year} {1991})}\BibitemShut {NoStop}%
\bibitem [{\citenamefont {Horn}\ and\ \citenamefont
  {Johnson}(2013)}]{horn2013}%
  \BibitemOpen
  \bibfield  {author} {\bibinfo {author} {\bibfnamefont {R.~A.}\ \bibnamefont
  {Horn}}\ and\ \bibinfo {author} {\bibfnamefont {C.~R.}\ \bibnamefont
  {Johnson}},\ }\href@noop {} {\emph {\bibinfo {title} {Matrix {{Analysis}}}}}\
  (\bibinfo  {publisher} {{Cambridge University Press}},\ \bibinfo {year}
  {2013})\BibitemShut {NoStop}%
\bibitem [{\citenamefont {Sottile}(2008)}]{sottile2008arXiv:math/0212044}%
  \BibitemOpen
  \bibfield  {author} {\bibinfo {author} {\bibfnamefont {F.}~\bibnamefont
  {Sottile}},\ }\bibfield  {title} {\bibinfo {title} {Toric ideals, real toric
  varieties, and the algebraic moment map},\ }\href@noop {} {\bibfield
  {journal} {\bibinfo  {journal} {arXiv:math/0212044}\ } (\bibinfo {year}
  {2008})},\ \Eprint {https://arxiv.org/abs/math/0212044} {arXiv:math/0212044}
  \BibitemShut {NoStop}%
\bibitem [{\citenamefont {Eisenbud}\ and\ \citenamefont
  {Sturmfels}(1996)}]{eisenbud1996DukeMath.J.}%
  \BibitemOpen
  \bibfield  {author} {\bibinfo {author} {\bibfnamefont {D.}~\bibnamefont
  {Eisenbud}}\ and\ \bibinfo {author} {\bibfnamefont {B.}~\bibnamefont
  {Sturmfels}},\ }\bibfield  {title} {\bibinfo {title} {Binomial ideals},\
  }\href {https://doi.org/10.1215/S0012-7094-96-08401-X} {\bibfield  {journal}
  {\bibinfo  {journal} {Duke Mathematical Journal}\ }\textbf {\bibinfo {volume}
  {84}},\ \bibinfo {pages} {1} (\bibinfo {year} {1996})}\BibitemShut {NoStop}%
\bibitem [{\citenamefont {Cox}\ \emph {et~al.}(2011)\citenamefont {Cox},
  \citenamefont {Little},\ and\ \citenamefont {Schenck}}]{cox2011}%
  \BibitemOpen
  \bibfield  {author} {\bibinfo {author} {\bibfnamefont {D.~A.}\ \bibnamefont
  {Cox}}, \bibinfo {author} {\bibfnamefont {J.~B.}\ \bibnamefont {Little}},\
  and\ \bibinfo {author} {\bibfnamefont {H.~K.}\ \bibnamefont {Schenck}},\
  }\href@noop {} {\emph {\bibinfo {title} {Toric {{Varieties}}}}}\ (\bibinfo
  {publisher} {{American Mathematical Soc.}},\ \bibinfo {year}
  {2011})\BibitemShut {NoStop}%
\bibitem [{\citenamefont {Callen}\ and\ \citenamefont
  {Callen}(1985)}]{callen1985}%
  \BibitemOpen
  \bibfield  {author} {\bibinfo {author} {\bibfnamefont {H.~B.}\ \bibnamefont
  {Callen}}\ and\ \bibinfo {author} {\bibfnamefont {H.~B.}\ \bibnamefont
  {Callen}},\ }\href@noop {} {\emph {\bibinfo {title} {Thermodynamics and an
  {{Introduction}} to {{Thermostatistics}}}}}\ (\bibinfo  {publisher}
  {{Wiley}},\ \bibinfo {year} {1985})\BibitemShut {NoStop}%
\bibitem [{\citenamefont {Pachter}\ and\ \citenamefont
  {Sturmfels}(2005)}]{pachter2005}%
  \BibitemOpen
  \bibinfo {editor} {\bibfnamefont {L.}~\bibnamefont {Pachter}}\ and\ \bibinfo
  {editor} {\bibfnamefont {B.}~\bibnamefont {Sturmfels}},\ eds.,\ \href
  {https://doi.org/10.1017/CBO9780511610684} {\emph {\bibinfo {title}
  {Algebraic {{Statistics}} for {{Computational Biology}}}}}\ (\bibinfo
  {publisher} {{Cambridge University Press}},\ \bibinfo {address}
  {{Cambridge}},\ \bibinfo {year} {2005})\BibitemShut {NoStop}%
\bibitem [{\citenamefont
  {Bregman}(1967)}]{bregman1967USSRComputationalMathematicsandMathematicalPhysics}%
  \BibitemOpen
  \bibfield  {author} {\bibinfo {author} {\bibfnamefont {L.~M.}\ \bibnamefont
  {Bregman}},\ }\bibfield  {title} {\bibinfo {title} {The relaxation method of
  finding the common point of convex sets and its application to the solution
  of problems in convex programming},\ }\href
  {https://doi.org/10.1016/0041-5553(67)90040-7} {\bibfield  {journal}
  {\bibinfo  {journal} USSR Computational Mathematics and Mathematical
  Physics\ }\textbf {\bibinfo {volume} {7}},\ \bibinfo {pages} {200} (\bibinfo
  {year} {1967})}\BibitemShut {NoStop}%
\bibitem [{\citenamefont {Shear}(1967)}]{shear1967JournalofTheoreticalBiology}%
  \BibitemOpen
  \bibfield  {author} {\bibinfo {author} {\bibfnamefont {D.}~\bibnamefont
  {Shear}},\ }\bibfield  {title} {\bibinfo {title} {An analog of the
  {{Boltzmann H}}-theorem (a {{Liapunov}} function) for systems of coupled
  chemical reactions},\ }\href {https://doi.org/10.1016/0022-5193(67)90005-7}
  {\bibfield  {journal} {\bibinfo  {journal} {Journal of Theoretical Biology}\
  }\textbf {\bibinfo {volume} {16}},\ \bibinfo {pages} {212} (\bibinfo {year}
  {1967})}\BibitemShut {NoStop}%
\bibitem [{\citenamefont
  {Higgins}(1968)}]{higgins1968JournalofTheoreticalBiology}%
  \BibitemOpen
  \bibfield  {author} {\bibinfo {author} {\bibfnamefont {J.}~\bibnamefont
  {Higgins}},\ }\bibfield  {title} {\bibinfo {title} {Some remarks on
  {{Shear}}'s {{Liapunov}} function for systems of chemical reactions},\ }\href
  {https://doi.org/10.1016/0022-5193(68)90117-3} {\bibfield  {journal}
  {\bibinfo  {journal} {Journal of Theoretical Biology}\ }\textbf {\bibinfo
  {volume} {21}},\ \bibinfo {pages} {293} (\bibinfo {year} {1968})}\BibitemShut
  {NoStop}%
\bibitem [{\citenamefont {Seifert}(2012)}]{seifert2012Rep.Prog.Phys.}%
  \BibitemOpen
  \bibfield  {author} {\bibinfo {author} {\bibfnamefont {U.}~\bibnamefont
  {Seifert}},\ }\bibfield  {title} {\bibinfo {title} {Stochastic
  thermodynamics, fluctuation theorems and molecular machines},\ }\href
  {https://doi.org/10.1088/0034-4885/75/12/126001} {\bibfield  {journal}
  {\bibinfo  {journal} {Reports on Progress in Physics}\ }\textbf {\bibinfo
  {volume} {75}},\ \bibinfo {pages} {126001} (\bibinfo {year}
  {2012})}\BibitemShut {NoStop}%
\bibitem [{\citenamefont {Ito}(2018)}]{ito2018Phys.Rev.Lett.}%
  \BibitemOpen
  \bibfield  {author} {\bibinfo {author} {\bibfnamefont {S.}~\bibnamefont
  {Ito}},\ }\bibfield  {title} {\bibinfo {title} {Stochastic {{Thermodynamic
  Interpretation}} of {{Information Geometry}}},\ }\href
  {https://doi.org/10.1103/PhysRevLett.121.030605} {\bibfield  {journal}
  {\bibinfo  {journal} {Physical Review Letters}\ }\textbf {\bibinfo {volume}
  {121}},\ \bibinfo {pages} {030605} (\bibinfo {year} {2018})}\BibitemShut
  {NoStop}%
\bibitem [{\citenamefont {Ohga}\ and\ \citenamefont
  {Ito}(2021{\natexlab{a}})}]{ohga2021ArXiv211211008Cond-Mat}%
  \BibitemOpen
  \bibfield  {author} {\bibinfo {author} {\bibfnamefont {N.}~\bibnamefont
  {Ohga}}\ and\ \bibinfo {author} {\bibfnamefont {S.}~\bibnamefont {Ito}},\
  }\bibfield  {title} {\bibinfo {title} {Information-geometric {{Legendre}}
  duality in stochastic thermodynamics},\ }\href@noop {} {\bibfield  {journal}
  {\bibinfo  {journal} {arXiv:2112.11008 [cond-mat]}\ } (\bibinfo {year}
  {2021}{\natexlab{a}})},\ \Eprint {https://arxiv.org/abs/2112.11008}
  {arXiv:2112.11008 [cond-mat]} \BibitemShut {NoStop}%
\bibitem [{\citenamefont {Adamer}\ and\ \citenamefont
  {Helmer}(2020)}]{adamer2020JMathChem}%
  \BibitemOpen
  \bibfield  {author} {\bibinfo {author} {\bibfnamefont {M.~F.}\ \bibnamefont
  {Adamer}}\ and\ \bibinfo {author} {\bibfnamefont {M.}~\bibnamefont
  {Helmer}},\ }\bibfield  {title} {\bibinfo {title} {Families of toric chemical
  reaction networks},\ }\href {https://doi.org/10.1007/s10910-020-01162-x}
  {\bibfield  {journal} {\bibinfo  {journal} {Journal of Mathematical
  Chemistry}\ }\textbf {\bibinfo {volume} {58}},\ \bibinfo {pages} {2061}
  (\bibinfo {year} {2020})}\BibitemShut {NoStop}%
\bibitem [{\citenamefont {Ay}\ \emph {et~al.}(2017)\citenamefont {Ay},
  \citenamefont {Jost}, \citenamefont {L{\^e}},\ and\ \citenamefont
  {Schwachh{\"o}fer}}]{ay2017}%
  \BibitemOpen
  \bibfield  {author} {\bibinfo {author} {\bibfnamefont {N.}~\bibnamefont
  {Ay}}, \bibinfo {author} {\bibfnamefont {J.}~\bibnamefont {Jost}}, \bibinfo
  {author} {\bibfnamefont {H.~V.}\ \bibnamefont {L{\^e}}},\ and\ \bibinfo
  {author} {\bibfnamefont {L.}~\bibnamefont {Schwachh{\"o}fer}},\ }\href@noop
  {} {\emph {\bibinfo {title} {Information {{Geometry}}}}}\ (\bibinfo
  {publisher} {{Springer}},\ \bibinfo {year} {2017})\BibitemShut {NoStop}%
\bibitem [{\citenamefont
  {Newton}(2012)}]{newton2012JournalofFunctionalAnalysis}%
  \BibitemOpen
  \bibfield  {author} {\bibinfo {author} {\bibfnamefont {N.~J.}\ \bibnamefont
  {Newton}},\ }\bibfield  {title} {\bibinfo {title} {An infinite-dimensional
  statistical manifold modelled on {{Hilbert}} space},\ }\href
  {https://doi.org/10.1016/j.jfa.2012.06.007} {\bibfield  {journal} {\bibinfo
  {journal} {Journal of Functional Analysis}\ }\textbf {\bibinfo {volume}
  {263}},\ \bibinfo {pages} {1661} (\bibinfo {year} {2012})}\BibitemShut
  {NoStop}%
\bibitem [{\citenamefont {Crooks}(2007)}]{crooks2007Phys.Rev.Lett.}%
  \BibitemOpen
  \bibfield  {author} {\bibinfo {author} {\bibfnamefont {G.~E.}\ \bibnamefont
  {Crooks}},\ }\bibfield  {title} {\bibinfo {title} {Measuring {{Thermodynamic
  Length}}},\ }\href {https://doi.org/10.1103/PhysRevLett.99.100602} {\bibfield
   {journal} {\bibinfo  {journal} {Physical Review Letters}\ }\textbf {\bibinfo
  {volume} {99}},\ \bibinfo {pages} {100602} (\bibinfo {year}
  {2007})}\BibitemShut {NoStop}%
\bibitem [{\citenamefont {Kolchinsky}\ and\ \citenamefont
  {Wolpert}(2021)}]{kolchinsky2021Phys.Rev.X}%
  \BibitemOpen
  \bibfield  {author} {\bibinfo {author} {\bibfnamefont {A.}~\bibnamefont
  {Kolchinsky}}\ and\ \bibinfo {author} {\bibfnamefont {D.~H.}\ \bibnamefont
  {Wolpert}},\ }\bibfield  {title} {\bibinfo {title} {Work, {{Entropy
  Production}}, and {{Thermodynamics}} of {{Information}} under {{Protocol
  Constraints}}},\ }\href {https://doi.org/10.1103/PhysRevX.11.041024}
  {\bibfield  {journal} {\bibinfo  {journal} {Physical Review X}\ }\textbf
  {\bibinfo {volume} {11}},\ \bibinfo {pages} {041024} (\bibinfo {year}
  {2021})}\BibitemShut {NoStop}%
\bibitem [{\citenamefont {Aris}(1965)}]{aris1965}%
  \BibitemOpen
  \bibfield  {author} {\bibinfo {author} {\bibfnamefont {R.}~\bibnamefont
  {Aris}},\ }\href@noop {} {\emph {\bibinfo {title} {Introduction to the
  {{Analysis}} of {{Chemical Reactors}}}}}\ (\bibinfo  {publisher}
  {{Prentice-Hall}},\ \bibinfo {year} {1965})\BibitemShut {NoStop}%
\bibitem [{\citenamefont {Qian}\ and\ \citenamefont {Ge}(2021)}]{qian2021}%
  \BibitemOpen
  \bibfield  {author} {\bibinfo {author} {\bibfnamefont {H.}~\bibnamefont
  {Qian}}\ and\ \bibinfo {author} {\bibfnamefont {H.}~\bibnamefont {Ge}},\
  }\href@noop {} {\emph {\bibinfo {title} {Stochastic {{Chemical Reaction
  Systems}} in {{Biology}}}}}\ (\bibinfo  {publisher} {{Springer International
  Publishing}},\ \bibinfo {year} {2021})\BibitemShut {NoStop}%
\bibitem [{\citenamefont {Ohga}\ and\ \citenamefont
  {Ito}(2021{\natexlab{b}})}]{ohga2021ArXiv211213813Cond-MatPhysicsphysicsa}%
  \BibitemOpen
  \bibfield  {author} {\bibinfo {author} {\bibfnamefont {N.}~\bibnamefont
  {Ohga}}\ and\ \bibinfo {author} {\bibfnamefont {S.}~\bibnamefont {Ito}},\
  }\bibfield  {title} {\bibinfo {title} {Information-geometric dual affine
  coordinate systems for chemical thermodynamics},\ }\href@noop {} {\bibfield
  {journal} {\bibinfo  {journal} {arXiv:2112.13813 [cond-mat,
  physics:physics]}\ } (\bibinfo {year} {2021}{\natexlab{b}})},\ \Eprint
  {https://arxiv.org/abs/2112.13813} {arXiv:2112.13813 [cond-mat,
  physics:physics]} \BibitemShut {NoStop}%
\end{thebibliography}%

\end{document}